\documentclass[a4paper,11pt]{article}
\usepackage{blindtext}
\usepackage{geometry}
\usepackage[affil-it]{authblk}

\usepackage{graphicx} 
\usepackage{subfig}
\usepackage{titlesec} 
\graphicspath{ {./images/} }

\usepackage[rightcaption]{sidecap}
\usepackage[utf8]{inputenc}
\usepackage{graphicx}
\usepackage{float} 
\usepackage{rotating} 
\usepackage{wrapfig}
\usepackage{amsmath}
\usepackage{textcomp}
\usepackage{tabularx}
\usepackage{hyperref}
\usepackage{placeins}
\usepackage{caption}
\usepackage{subcaption}
\usepackage{setspace} 

\title{\textbf{On the flow topology of swirl jets upon impingement}}

\author[1]{Premchand V Chandra}

\author[2,\S]{Pradip Dutta}

\affil[1,2]{Indian Institute of Science, Bangalore, India}

\affil[ \S]{Author Correspondence:- pradip@iisc.ac.in; premchandv@iisc.ac.in}

\begin{document}
\maketitle
\begin{abstract}
    Jet impingement enhances heat transfer and is characterized by the complex flow patterns formed when a jet impacts a plate aligned normal to it. While traditional round jet impingement has been extensively studied to comprehend flow and associated heat transfer, there is still room for research when it comes to the investigation of flow structures in swirl jet impingement. This paper focuses on the flow topology of swirl jets generated by a $45^\circ$ axial vane swirler, impinging on a flat plate studied at dimensionless jet-plate distances (H/D = 1 – 4) and Reynolds numbers (Re = 16600, and 23000). The flow structures, the mean velocity components, and the turbulence characteristics using a 2D- Particle Image Velocimetry (PIV) experiments at the front $(r- z)$ and top $(r- \theta)$ planes of impingement are presented. Furthermore, results from the 3D numerical simulations are presented to support the results where PIV study had experimental limitations. The effect of impingement distance or jet -plate distance $(H/D)$ on the mean flow properties and turbulence parameters are discussed. A Proper Orthogonal Decomposition (POD) analysis has been performed to understand the dominant coherent structures at different cases of impingement distance (H/D). We show that the turbulence parameters are more pronounced at smaller jet-plate distances $(H/D \leq 2)$ which could reason out the enhanced heat transfer for these jets.

\end{abstract}
\section{Introduction}
Jet impingement is a widely adopted technique in engineering and industrial processes due to its high efficiency in transferring heat and mass. Applications range from cooling of turbine blades, electronic chips, and nuclear reactors to drying processes, surface cleaning, and material processing. Additionally, jet impingement finds significant roles in momentum-dominated processes such as Rocket propulsion, fuel injection, fire suppression, shielded arc welding, cutting, and paint spraying (Martin 1977, and Goldstein et al., 1982). Among the different types of jets, round jets have been extensively investigated due to their simple geometry and well-defined flow characteristics. The flow topology of round jets, upon impingement, typically comprises of the free jet region, followed by the stagnation region and the radial wall jet region. Donalson et.al (1971), performed an experimental study to understand the behaviour of normally and obliquely impinging jets. They detailed the mean flow properties, turbulent structures and the radial velocity gradient to calculate the heat transfer at the stagnation point, along with surface pressure measurements and wall jets. Experimental works by Cooper et al. (1993), detail both flow and heat transfer aspects of the fully developed round jet at Re=23000 and H/D =2 and 6, to provide hydrodynamic data for the same conditions of Baughn and Shimizu (1989) experiments. Their hot-wire measurements reported the mean velocity profile and Reynolds-stress components in the vicinity of the impinging plate surface, besides reporting the Nusselt number data.

 Further experimental and numerical studies have revealed that round jets exhibit self-similar behaviour downstream, with strong turbulence in the wall jet region (Hassan, 2019). In turbulent round jets, the mean flow field exhibits axi-symmetry with significant Reynolds stress distributions in the shear layers due to velocity gradients between the jet core and the ambient fluid. Studies by Phares et al. (2000) put forth analytical solutions to the stream-vorticity equation for 2D impingement flow and axisymmetric impingement flow with arbitrary inlet velocity profile. The round jet impingement literatures are rich interms of experimental, computational and analytical investigations. These investigations have laid the foundation for understanding round jets topology, flow field and heat transfer aspects; however, their limited mixing capabilities have motivated researchers to explore swirl jets, where flow characteristics are fundamentally different. 
 
 Swirl jets have emerged as a more versatile alternative owing to their ability to enhance mixing, radial spreading, and turbulence control. Swirl jets are generated by introducing a tangential component of velocity, which produces rotational motion in addition to the axial flow. This swirl motion is characterised by the swirl number (S), defined as the ratio of angular momentum to axial momentum flux. Swirl jets exhibit highly complex and three-dimensional flow structures compared to conventional round jets, leading to vortex breakdown, recirculation zones, and enhanced mixing. At the nozzle exit, swirl jets display a central toroidal recirculation zone (CTRZ) or vortex breakdown due to the balance between centrifugal and pressure forces (Gupta et al., 1984). These recirculation zones are strongly dependent on the swirl number and Reynolds number. For low swirl intensities, the flow remains primarily axial, while at higher swirl numbers, the flow transitions to a highly rotational regime with significant radial spreading.
 
 Upon impingement, the swirl jet's topology undergoes further transformation. Unlike round jets, swirl jets produce a broader stagnation region due to the radial outward flow driven by centrifugal forces. The swirling motion generates secondary vortices and shear layers, enhancing turbulent mixing and momentum transfer along the surface (Ekkad and Huang, 2002). Swirl jets often create large recirculating regions downstream of the impingement surface, which impact heat transfer and flow uniformity. 
 
 The turbulent characteristics of swirl jets upon impingement are notably distinct from round jets. Large Eddy Simulations (LES) and Direct Numerical Simulations (DNS) by Ashforth-Frost et al. (1996) and Patankar et al. (1995) revealed intense shear stress and Reynolds stresses near the impingement surface due to vortex breakdown. Swirl jets also display unsteady flow patterns, including secondary vortices and periodic oscillations, which influence the stagnation region's heat transfer distribution.
Studies have extended this understanding through experimental investigations using Particle Image Velocimetry (PIV) and Laser Doppler Anemometry (LDA). These techniques have provided detailed measurements of the velocity field, turbulence intensity, and vortex dynamics of impinging swirl jets . However, the complex interaction between the swirl intensity, Reynolds number, and nozzle-to-surface spacing remains underexplored, particularly for the near-wall flow structures and shear stress distributions. 
While substantial work has been conducted on round jet impingement, studies on swirl jets are still limited, especially concerning the flow topology and turbulence characteristics near the impingement surface. Existing studies primarily focus on heat transfer performance, with insufficient attention given to the detailed flow dynamics, such as recirculation zones, vortex breakdown, and secondary flow formation. Furthermore, experimental validation of numerical predictions for swirl jet impingement is scarce, particularly for intermediate and high swirl intensities.

This study aims to address these gaps by investigating the flow topology of swirl jets upon impingement, with more emphasis given on the parameter jet-to-plate distance $(H/D)$. Using Particle Image Velocimetry (PIV) and 3D-simulations, we analyze the near-wall flow structures, vortex dynamics, and turbulence characteristics to provide a comprehensive understanding of the impingement process. The specific objectives of this work are to characterise the mean flow field and turbulent structures of impinging swirl jets and to investigate the influence of jet-to-surface spacing on the flow topology and shear stress distribution, especially at lower impingement distances between $H = 1 – 4D$. The scope of this work is to elucidate the importance of lower jet-plate distance $( H/D\leq2)$ in the swirl jet impingement, its significance of enhanced turbulence and mean flow characteristics, which could explain the enhanced heat transfer by these jets.
\section{Methodology}
Here, we use 2D-PIV experiments to study the flow fields at the orthogonal front $(r-z)$ plane and at top $(r- \theta)$ plane orientations. Also, 3D simulation using commercial software ANSYS-Fluent has been used where experiments had practical limitations to delineate the flow topology, average flow and turbulent statistics (free jets results are detailed in the Appendix). The important parameters studied are the jet-plate distance $(H/D = 1.5, 2, 3, \& 4)$ and the inlet Reynolds number in the turbulent regime, especially at $(Re = 16600,\& 23000)$, of which the $Re = 23000$ has an important significance in the jet impingement literature. This section discusses the details of the test section, the vane swirler geometry chosen for the present study, description of the PIV techniques, computational methods, experimental measurements methods, and data analysis.
\subsection{Swirler geometry for the experimental study}
In the present study, we choose a geometrical vane swirler adapted from the one used for flame stabilisation and thermo-acoustic instability studies. The computational validation of the model using the above geometrical vane swirler geometry is already shown in our earlier article (Chandra PV et al. 2023), which validates the axial and azimuthal (tangential) components of velocity . A part of the above work by Chandra et al. investigated the flow field and heat transfer characteristics using the vane swirler, but was limited to a computational study. The present work is a continuation of the above work to delineate the flow topologies using flow visualization PIV experiments. The same swirler geometry is used for 3D-computational studies, where PIV experiments had limitations.

\begin{figure}[h!]
\centering
\includegraphics[scale = 0.9]{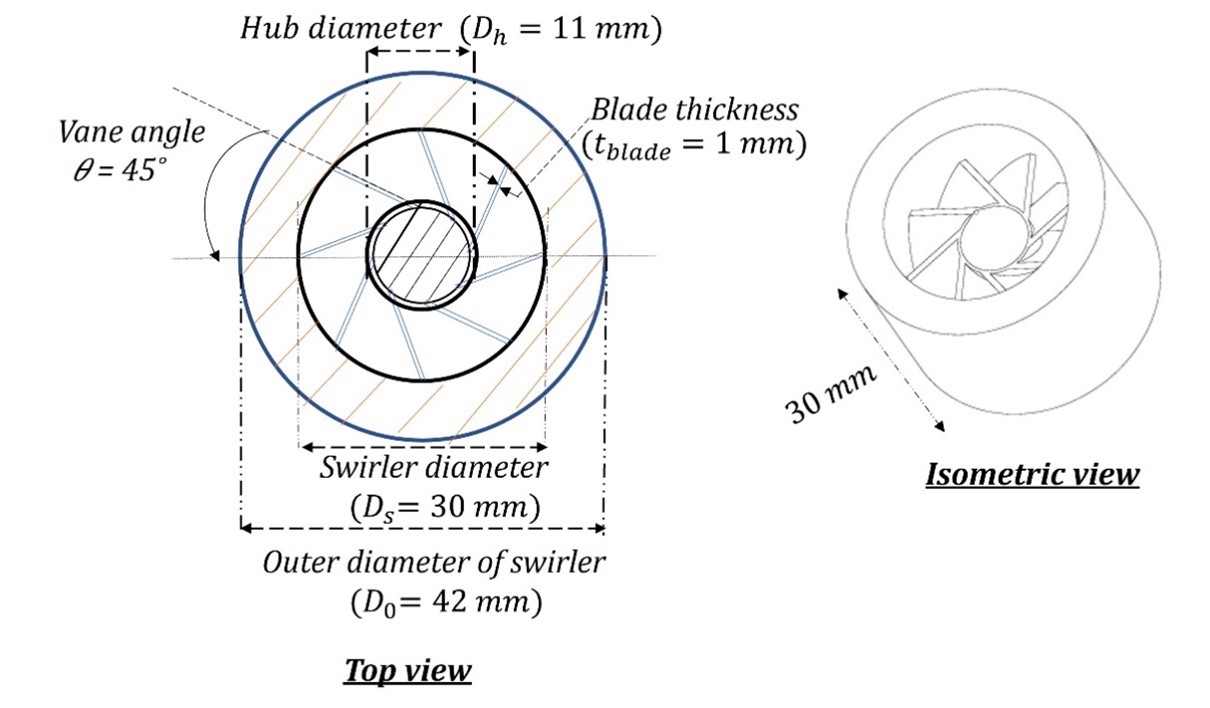}
\caption{The vane swirler chosen for the present study. The left figure shows the top view of the swirler, which shows the geometrical details, and the right figure is the isometric view.}
\label{fig:Fig1}
\end{figure}

Figure 2 shows the geometrical details of the vane swirler, which has an inner diameter ID = 30mm, containing eight vanes inclined at an angle $(\theta = 45 ^{\circ})$ that corresponds to geometrical Swirl number (S) of 0.7 calculated from the following relation (Liley),
\begin{equation}\label{eq:1}
    S = \dfrac{2}{3} \left[ \dfrac{1 - \left(\dfrac{D_{h}}{D_{o}}\right)^{3}} {1 - \left(\dfrac{D_{h}}{D_{o}}\right)^{2}} \right] \tan \theta
\end{equation}
where $\dfrac{D_h}{D_o}$ is the ratio of hub diameter to the outer diameter of the swirler geometry and $\theta$ is the blade or vane angle in degrees. Owing to the blockage by the hub of the swirler, which holds the blades, the Reynolds number $(Re)$ calculated for this case is given by,
\begin{equation}\label{eq:2}
    Re = \dfrac{\rho u (D_{s} - D_{h})}{\mu} 
\end{equation}
where $D_s$ is the diameter of the swirler (equal to jet diameter D = 30 mm), and $D_h $= 11 mm is the hub diameter. 
\subsection{Experimental test section}
The test section consists of a swirler jet assembly shown on the right side of Fig.2, which has a single axial vane swirler whose details are given in Fig.1. An acrylic plate (impingement plate) was kept perpendicular to the jet assembly. The plate is mounted on a rack and pinion adjustable mechanism to move up and down vertically so that a variable jet-plate distance (H/D) can be realized. For the non-impingement (free) jet case PIV experiments are performed without the acrylic impingement plate. Pressurized air along with the olive oil tracer or seeding particles (for PIV experiments) is supplied to the swirler jet assembly through a tube of 10mm ID and length of 50 times the diameter i.e. 500mm. Air exits through the swirler exit, whose ID is 30mm, and expands in the open domain (unconfined or submerged swirl jet) normal to the impingement plate. The further details of the PIV are detailed in the following section 2.3 and Fig 3.
\begin{figure}[h!]
\centering
\includegraphics[scale = 0.9]{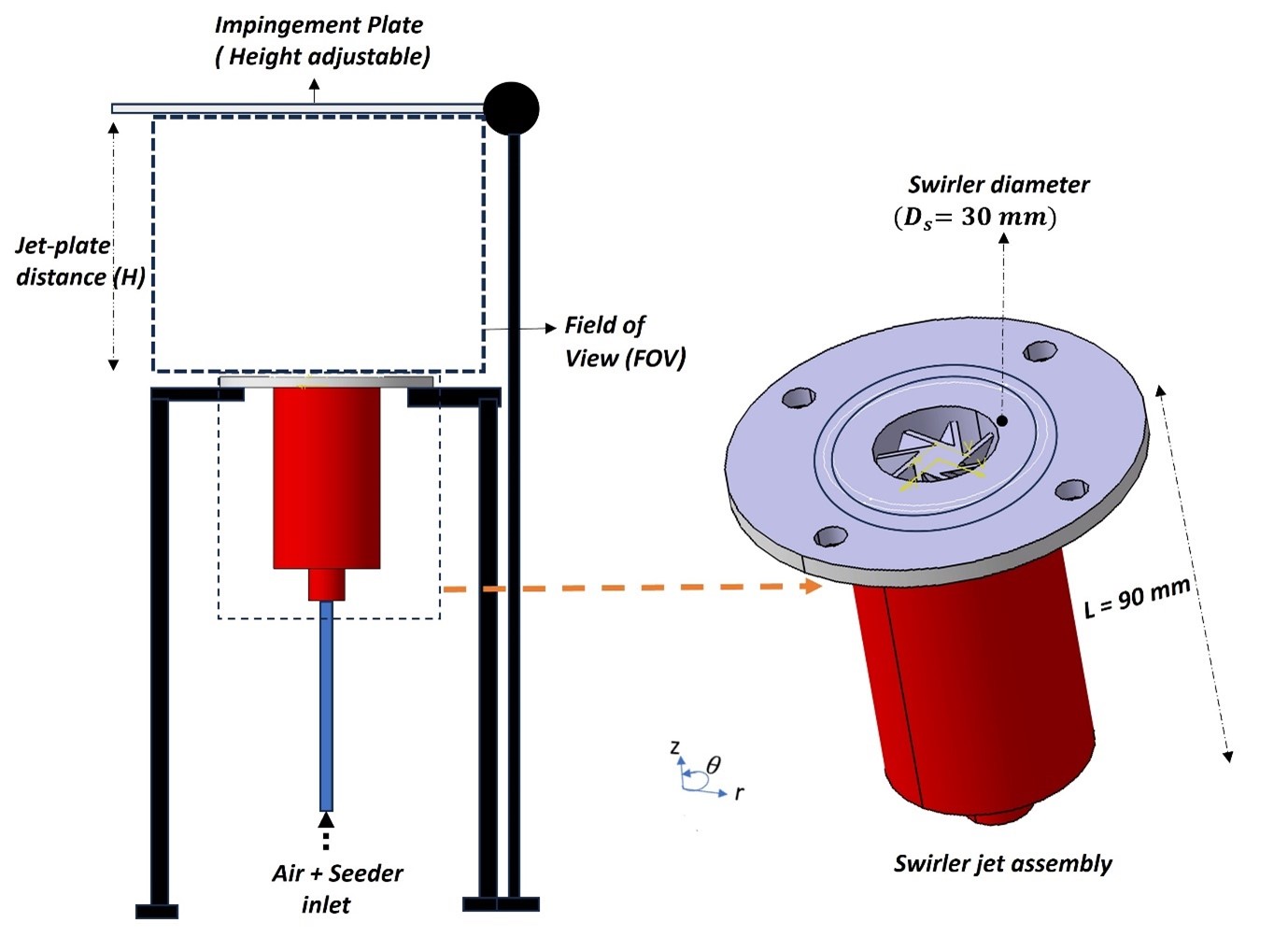}
\caption{Schematic of the experimental test section, which has a swirler jet assembly inside which the vane swirler is fixed.}
\label{fig:Fig2}
\end{figure}
\subsection{Particle Image Velocimetry (PIV) experiment}
Particle image velocimetry (PIV) flow visualization experiments are performed using a Nd:YAG (Photonics) dual pulse laser having 200mJ of energy, pulse width of 10ns and frequency of 10Hz. A laser sheet which illuminates the measurement plane (field of view) was formed using a sheet optics of 10mm focal length. Olive oil is used as the seeding particles after atomization into finite droplets approximately to $1\mu m$ using an atomizer. The PIV images are captured using a high-resolution charge-coupled device (CCD) camera (Imager SX, 4 Mega pixel) at a frame rate of 10Hz double exposure mode that allowed recording of two consecutive images. 

Steady flow of air is supplied form a compressed air facility maintained at 20 bar pressure. ALICAT mass flow controllers (MFC) are used for metering and controlling the air and seeding particles flow rates. The uncertainty of the MFC is reported to be ± 5 LPM of air for the range 100-700 LPM of the current experiments. Air at controlled flow rates corresponding to a particular inlet Reynolds no. (Re) enters the swirler jet assembly and exits through the swirler towards the impingement plate set at a desired jet-plate distance (H/D).

 A Programmable Timing Unit (PTU) simultaneously triggers the laser and camera which allows recording double images at 10Hz repetition rate of the laser. A total of 500 images were recorded for 50s at 10Hz for each case of the experiments. Based on the pixel shift of the particles in the interrogation window between the frames, the pulse interval (dt) was chosen between 26 – 42 $\mu$S. Lavision’s Davis 8.4 software was used for recording and post-processing of the data. The cross-correlation technique of the 2D images gives the velocity distribution or flow field from the displacement of the particles. For the cross correlation, the interrogation region was maintained at 64 x 64 pixels.
 
 \begin{figure}[h!]
\centering
\includegraphics[scale = 0.9]{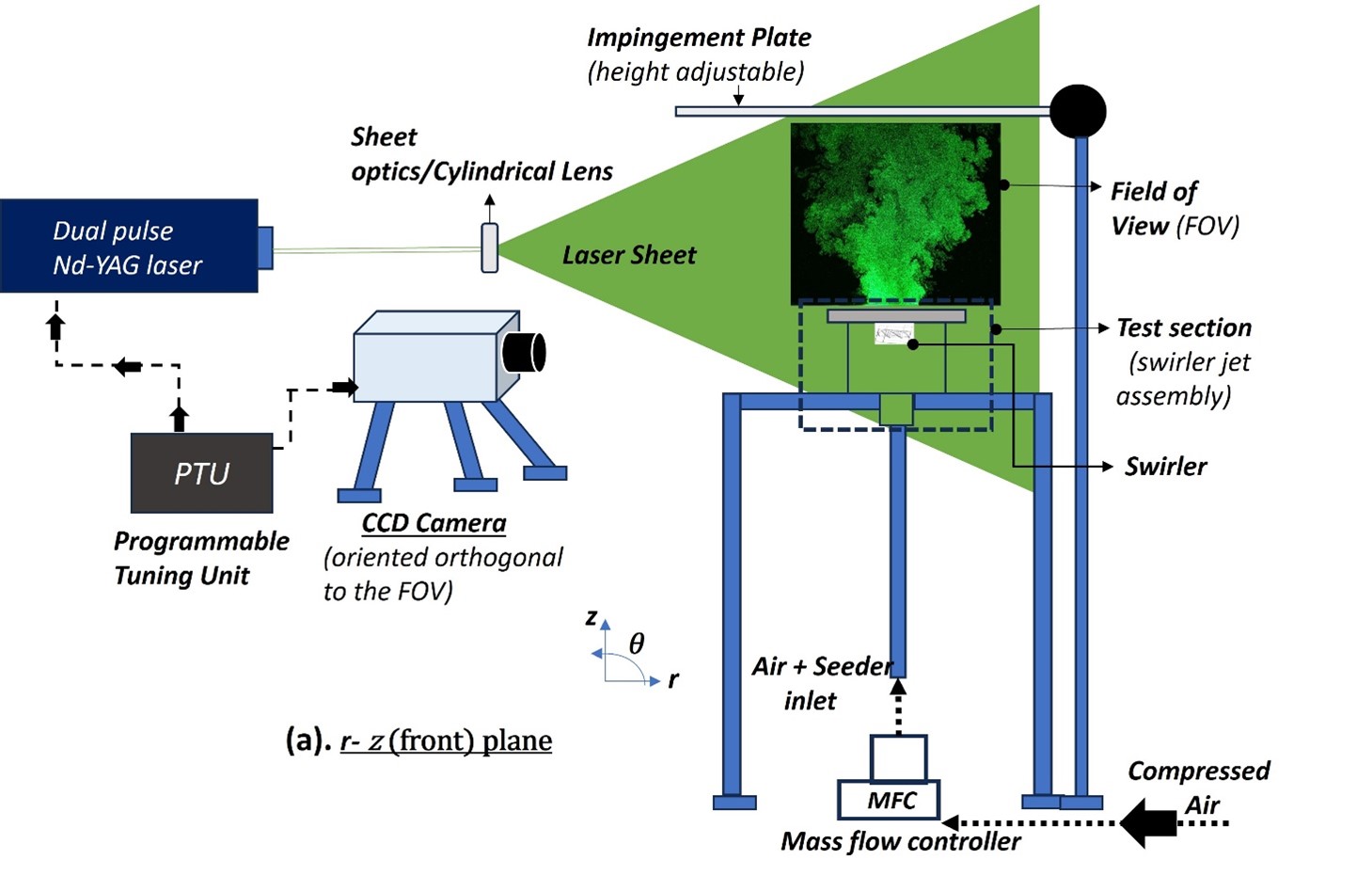}
\includegraphics[scale = 0.9]{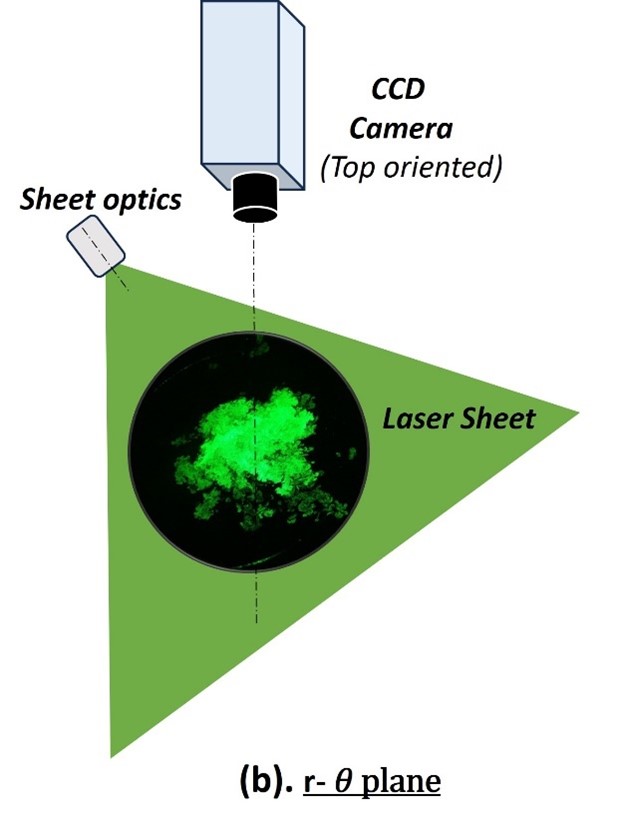}
\caption{Schematic of the PIV flow visualization setup (a). Shows the orientation of the setup for capturing the mid-longitudinal front plane (r-z) (b). Shows the illustration of illuminated laser sheet at the top $(r-\theta)$ plane and the orthogonally oriented camera.}
\label{fig:Fig3}
\end{figure}

Following is the illustration of the raw PIV images along the mid-longitudinal (or) front plane $(r- z)$ and top plane $(r-\theta)$ at z= 45 mm for the Reynolds number Re = 23000. The streamline from the average velocity is superimposed on the raw image in Fig. (c.).
\begin{figure}[h!]
\centering
\includegraphics[scale = 0.9]{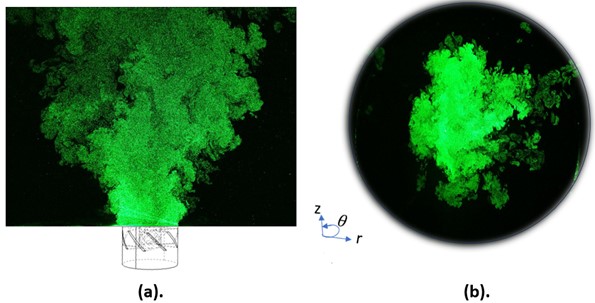}
\includegraphics[scale = 0.9]{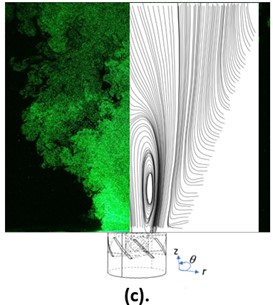}
\caption{Illustration of the 2D PIV flow field for a 450 vane swirler (Re=16600) from the experimental raw/instantaneous images at (a). Front (or) mid-longitudinal plane $(r-z)$ plane, and (b). Top $(r-\theta)$ plane at z = 45mm (c). Time-averaged streamlines (half-superimposed) on the front $(r-z)$ plane image.}
\label{fig:Fig3}
\end{figure}
\subsection{Computational methods and validation}
The Reynolds averaged continuity, momentum, and energy equations in coordinate-independent tensorial form are,\\
Continuity:
\begin{equation}\label{eq:1}
\dfrac{\partial \overline{u_{i}}}{\partial x_{i}}=0
\end{equation}
The momentum equation$(RANS)$,
\begin{equation}\label{eq:2}
    \rho \dfrac{\partial \overline{u_{i}}}{\partial t} + \rho \overline{u_{j}} \dfrac{\partial \overline{u_{i}}}{\partial \overline{x_{j}}} = - \dfrac{\partial \overline{p}}{\partial x_{i}} + \dfrac{\partial}{\partial x_{j}}(2\mu \overline{S_{ij}} - \rho \overline{u_{i}' u_{j}'})
\end{equation}
Where $\overline{u_{i}}$, $x_{i}$, $x_{j}$,$t$, $\overline{p}$, $\rho$, and $\mu$ are the mean velocity, spatial coordinates (in tensorial form with $i,j$ as the indices), temporal coordinate, mean pressure, density, and dynamic viscosity of air respectively. 
The mean strain rate is given by:
\begin{equation}\label{eq:5}
    \overline S_{ij} = \dfrac{1}{2} \left(\dfrac{\partial u_{i}}{\partial x_{j}} + \dfrac{\partial u_{j}}{\partial x_{i}} \right)
\end{equation}
The linear eddy viscosity model defines the Reynold’s stress which is given by,
\begin{equation}\label{eq:6}
    -\rho \overline {u_{i}' u_{j}'} = 2 \mu_{t} \overline{ \delta_{ij}} - \dfrac{2}{3} \rho k \delta_{ij}
\end{equation}

where $k$ is the turbulent kinetic energy, $\mu_{t}$ is the turbulent or eddy viscosity, $\delta_{ij}$ is the Kronecker delta, and $Pr_{t}$ is the turbulent Prandtl number. For the above RANS equations, modelling the Reynolds stress or turbulent stress term
The three-dimensional problem is solved for continuity, momentum, and energy equations in $r$, $\theta$, and $z$ directions along with transport equations for $ v-2-f$ model and Transition $ k-k_{l} - \omega$ turbulence models of RANS. The computational domain and the grid independence study have already been presented in our erstwhile work [Chandra et al.]. Here, we present the computational results of swirl flow velocity profiles computed using the v-2-f model, which are compared and validated with the PIV experimental data of the present work. 
\subsection{Validation plots}
The validation plots for axial and azimuthal (tangential) velocities for the free swirl jet generated by a $45^{\circ}$ vane swirler with the present PIV experiment are shown below in Figs . 5 \& 6. The axial velocities at two axial locations from the jet exit (z = 20mm, and 40mm) closely matches with the PIV results from the r-z mid-longitudinal plane experimental data.

\begin{figure}[h!]
\centering
\includegraphics[scale = 1.1]{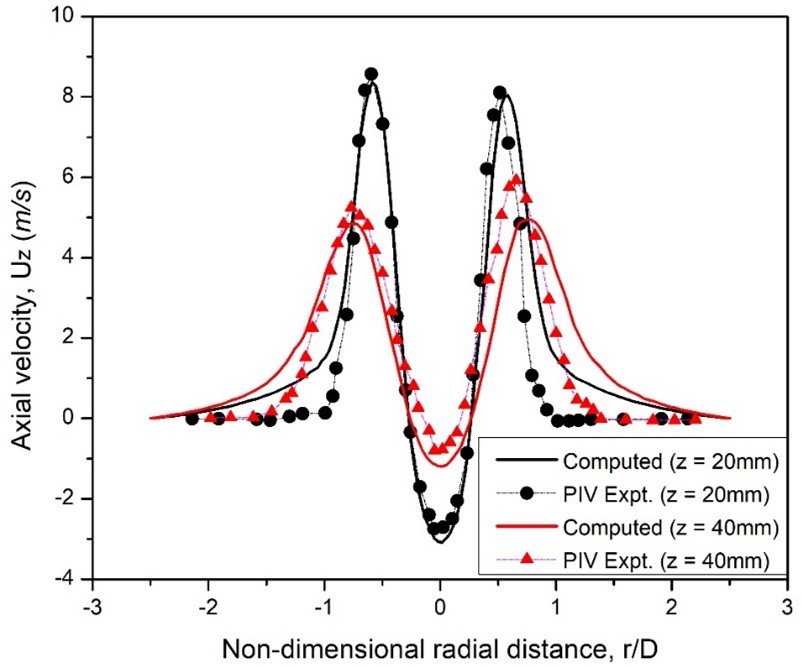}
\caption{Computed axial velocity at z = 20mm and 40mm validated with present PIV experiments at Re=16600 for the swirl jet by $45^\circ$ vane swirler (free jet).}
\label{fig:Fig5}
\end{figure}

\begin{figure}[h!]
\centering
\includegraphics[scale = 1.1]{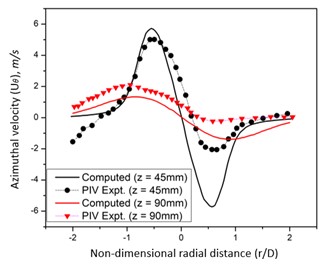}
\caption{Computed azimuthal velocity at $r- \theta$ planes z = 45mm and 90mm validated with present PIV experiments at Re=16600 for the swirl jet by $45^\circ$  vane swirler (free jet).}
\label{fig:Fig6}
\end{figure}

\begin{figure}[h!]
\centering
\includegraphics[scale = 0.9]{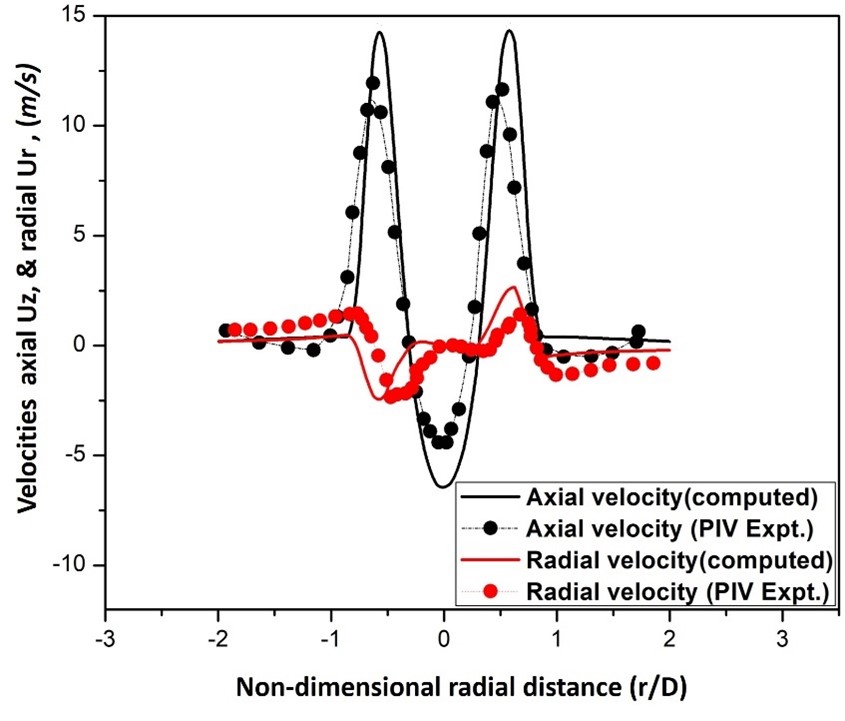}
\caption{Computed axial and radial velocity at z = 20mm for the impingement case validated with present PIV experiments at Re=16600 for the swirl jet by 450 vane swirler.}
\label{fig:Fig7}
\end{figure}

\begin{figure}[h!]
\centering
\includegraphics[scale = 0.9]{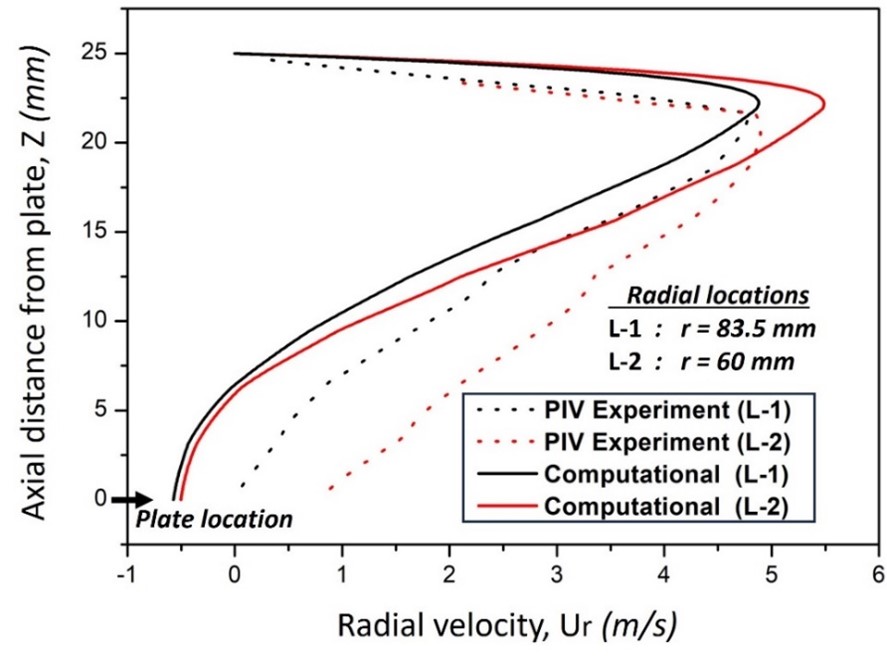}
\caption{Validation plot of present PIV experiments and computations for the radial velocity (near-field to the plate, upon impingement) at two radial locations in the wall jet (L-1 and L-2 whose radial locations are at r = 83.5 mm and r = 60 mm) for the case H/D = 3 at Re = 23000.}
\label{fig:Fig8}
\end{figure} 
The Fig.6. shows an asymmetry in the experimental results for the azimuthal velocity (from $r/D > 0$) which can be due to notable decay of swirl flows at z = 45mm and 90mm as literature suggests [W.G. Ross] that decay in swirl flows is highly desirable beyond z = 20 mm. Owing to experimental constraints, $  r-\theta$ planes, PIV data could not be acquired at such lower heights. Consequently, the validation plots also have notable mismatch between computed and experimental azimuthal velocity beyond $r/D > 0$ at these axial locations z = 45mm and 90mm. The asymmetry in the swirling jets is detailed further in the Appendix C which also describes the measure of asymmetry in the velocity for the decaying swirl jet at both impinging and non-impinging conditions.

The validation plot for the impingement case is shown in the below fig.7. The axial and radial velocities at z = 20mm are shown. For azimuthal velocity, PIV experiment cannot be performed for the impingement case as it requires overlap of the laser sheet with the acrylic impingement plate which results in scattering. The above event also imposes difficulty in capturing the top plane (impingement plane) field of view (FOV) with a CCD camera. However, 3D computational simulations help in plotting the azimuthal velocity and understanding flow structures for the impingement cases, which will be detailed in section 3.

The swirl jet upon impingement on the plate has the radial and azimuthal velocity at the near wall. These components of velocity form the wall jet which plays a significant role in convective transport properties. The validation plot for radial velocity at the near wall is shown in the below fig.8.
\newpage
\section{Results and Discussions}
The section presents the flow topology, the mean flow properties and turbulence statistics from the experiments and computations for the impingement case. A POD study for impinging jets at different jet-to-plate distances has also been discussed.
\subsection{Flow topology}

The flow field obtained from the PIV flow visualization experiments for the $45^\circ$ swirl jet for the impingement cases is discussed here. The flow topology from the 3D simulations is also presented. The freely swirling jet topology without impingement is detailed in Appendices A and B, which show the flow topology from PIV experiments and computations, respectively. 

\begin{figure}[h!]
\centering
\includegraphics[scale = 0.9]{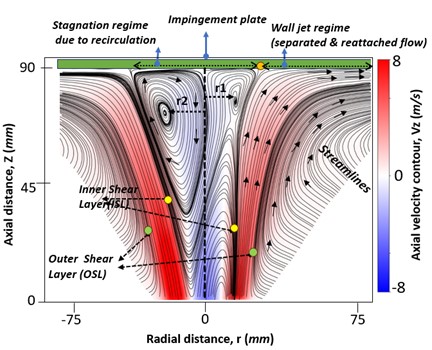}
\caption{Streamline and velocity contour (axial component) plot along the front plane (r-z) for a  $45^\circ$  swirl jet corresponding to Reynolds no. (Re = 16600) from the PIV experiment (a). Free jet (non-impingement) (b). Impingement case at H/D = 3 or H = 90 mm.}
\label{fig:Fig9}
\end{figure}

Unlike non-impinging jet cases, PIV experiments are not possible for the impinging cases as stated earlier. Hence, flow structures from the 3D simulations are presented in the following figure 10, which delineates the flow topology at multiple $(r-\theta)$ top planes.

\begin{figure}[h!]
\centering
\includegraphics[scale = 0.9]{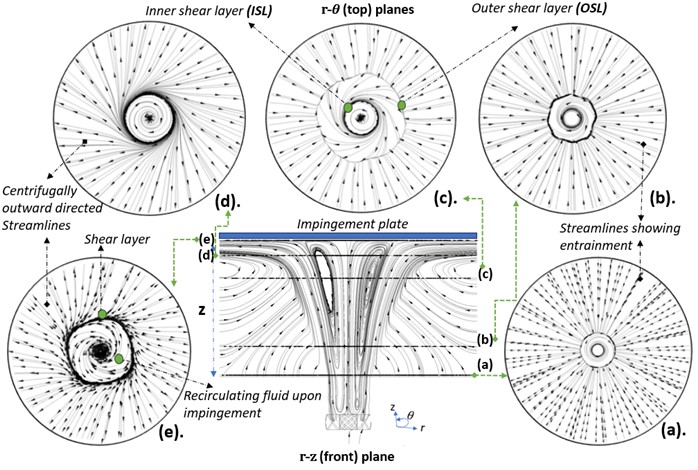}
\caption{Computed streamlines jet at meridional front and different top $r-\theta$ planes for the $45^\circ$ vane swirler at Re = 23000 (a). z = 0 mm (jet exit plane), (b). z = 20 mm, (c). z = 65 mm, (d). z = 80 mm, and (e). z = 90 (Impingement plane).}
\label{fig:Fig10}
\end{figure}
The evolution of the circular shear layers ISL and OSL can be appreciated from Fig. 10 (b) – (c). In the figures (a), (b) and (c), the radially inward streamlines from the periphery to the shear layer indicate the air entrainment from the surrounding. Clearly near the impingement plate cases (Fig.10. (d) and (e)). The streamlines are directed centrifugally outward, forming a single shear layer which encompasses a vortex motion, and a concentrated core stagnation region.
This aligns with a similar study and reasoning of our erstwhile work [Chandra et al.]
\begin{figure}[h!]
\centering
\includegraphics[scale = 0.9]{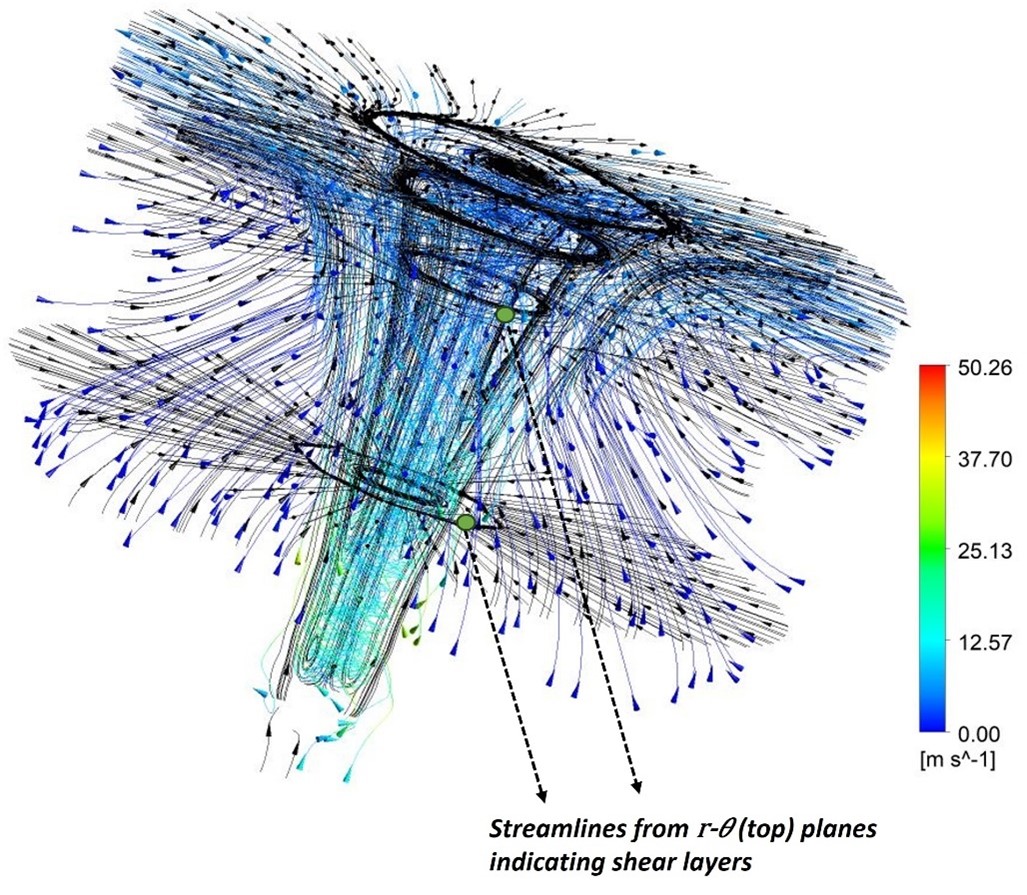}
\caption{Computed 3D streamlines for the Impingement case. The front plane is overlapped with the top $(r-\theta)$ planes at different axial locations z = 0mm (jet exit dump plane), 20mm, 65mm, 80mm, and 90mm.}
\label{fig:Fig11}
\end{figure}

Figure 11 shows the superimposed front $(r-z)$ and top $(r-\theta)$ plane streamlines from the 3D simulation results, which ascertain the entrainment, the evolution of shear layers and the radially outward jet motion after impingement with a central vortex core, which may contribute to a wider stagnation zone. 

\subsection{Mean flow Properties}
The swirl jet of diameter D = 30mm generated by a $45^\circ$ vane swirler at a range of Reynolds numbers (Re = 5000 – 23000) is studied at jet-plate impinging distances $(H/D = 1.5, 2, 3,\& 4)$ using a PIV experiment. Important results from the Re = 23000 case, which has an important significance in jet impingement literature and additionally at Re=16600 are discussed.

\subsubsection{Impinging jet mean velocity components}
The axial velocity profile tends towards a symmetric nature as the axial distance increases. Here we can see that at z = 20mm the velocity profile is asymmetric (in magnitude). As the axial distance increases gradually between z = 40 – 85 mm, the velocity plots tend towards a symmetric profile and at the impingement plate (z = 90mm), the velocity profile is flat, marking zero velocity magnitude due to obstruction of the plate. 
\begin{figure}[h!]
\centering
\includegraphics[scale = 0.95]{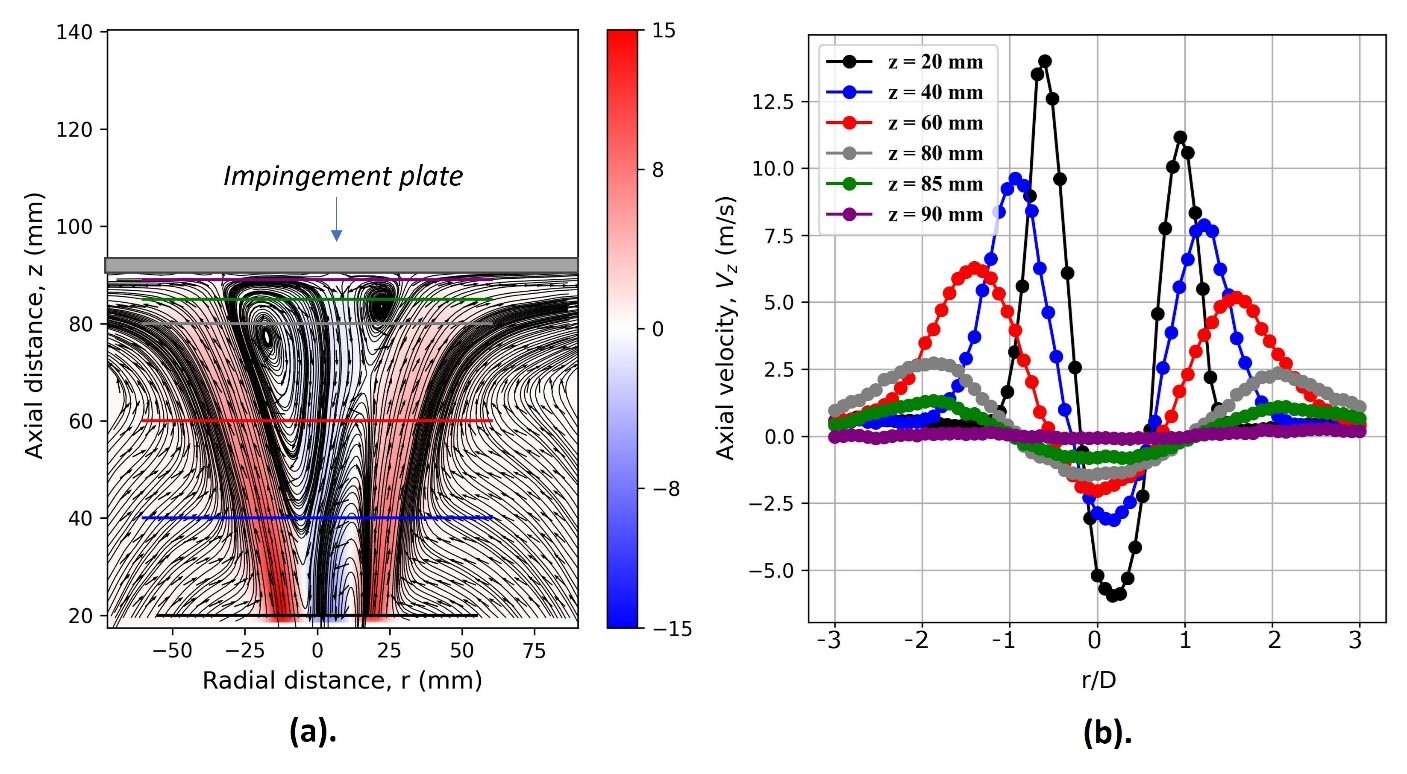}
\caption{Axial velocity components from the (r-z) plane for a 450 swirler impinging swirl jet at Re = 23000 (a). Streamlines and axial velocity contour plots (b). Axial velocity profiles at different axial locations (z = 20, 40, 60, and 80 mm) indicate their decay characteristics.}
\label{fig:Fig12}
\end{figure}

\begin{figure}[h!]
\centering
\includegraphics[scale = 0.95]{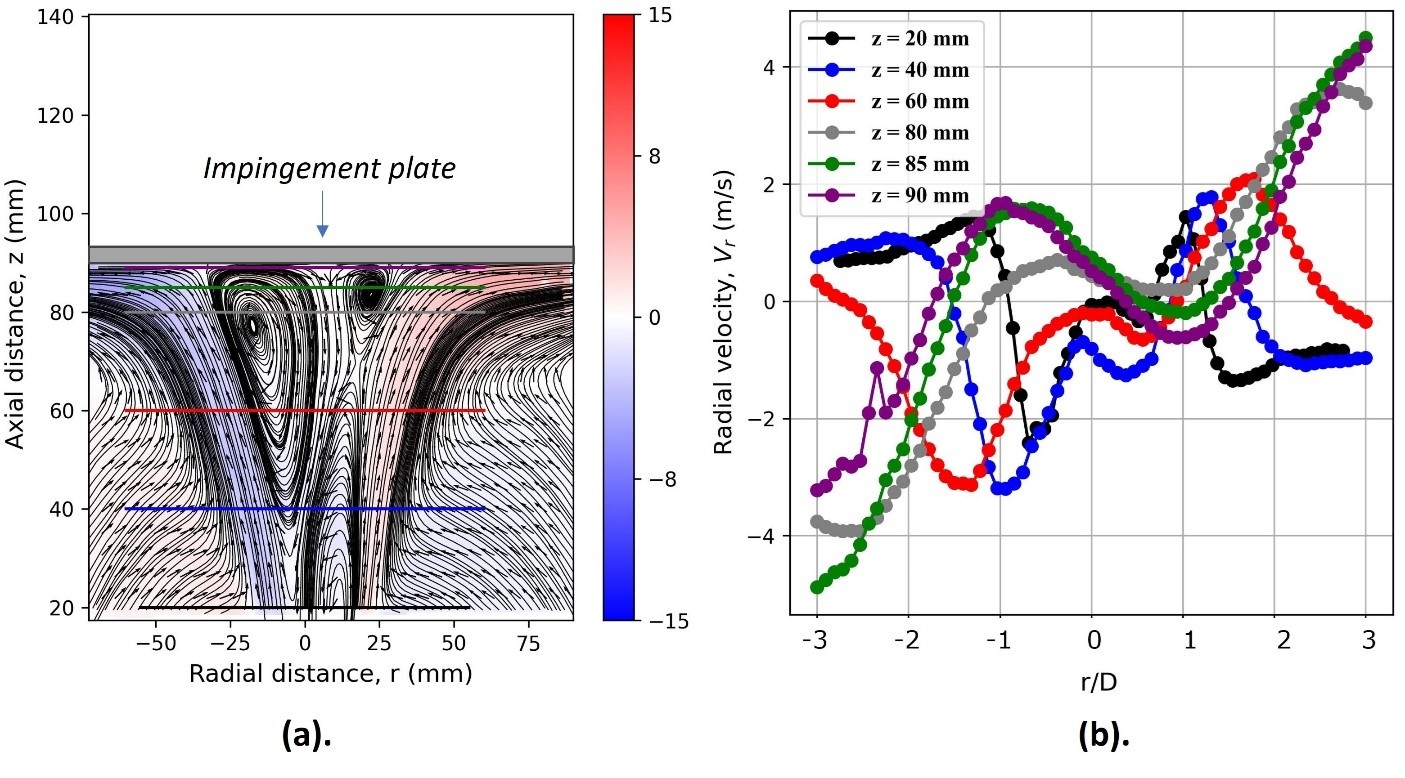}
\caption{Radial velocity components from the front r-z plane for a 450 swirler free jet (impingement) at Re = 23000 (a). Streamlines and radial velocity contour plots (b). Radial velocity profiles at different axial locations (z = 20, 40, 60, and 80 mm) for the impingement case indicate its growth characteristics.}
\label{fig:Fig13}
\end{figure}
From the comparative observations, it can be inferred that there is more organization of flow structures near the entrainment region from the surrounding to the jet core for the impingement cases when compared to freely swirling jet cases. The axial velocity gradually decreases towards the plate, which is compensated by increased radial velocity components. There is also a distinct organisation of radial velocity profiles with the presence of an impingement plate when compared to a non-impingement case, whose velocity profile appears to fluctuate.
\subsubsection{Near wall radial velocity profiles upon impingement}
From the point of jet impingement heat transfer analysis, the radial velocity, which facilitates the convective heat transfer along the impingement and wall jet regimes of the plate, is important, hence forming a substantial interest in this work. Figure 8(a). shows the details of the probed vertical lines L-1 to L-5, each of 25 mm in length from the impingement plate. L-1 is at the wall jet exit, L-2 is at the wall jet regime, L-3 and L-4 are at the impingement regime, followed by L-5 at the stagnation. The radial velocity profiles along 5 different locations (L-1 to L-5) are plotted as shown in Figure 8 (b) for H/D = 3 at Re = 23000.

It is observed that the radial velocity profile with a maximum velocity component can be seen along line L-2, which falls under the wall jet regime, followed by L-1 (wall jet exit). At L-5, which corresponds to the stagnation regime, the flow diverts radially, making the radial velocity close to zero at a stagnation point. A similar velocity profile is shown for a lower impinging distance, H/D = 2. 

\begin{figure}[h!]
\centering
\includegraphics[scale = 0.9]{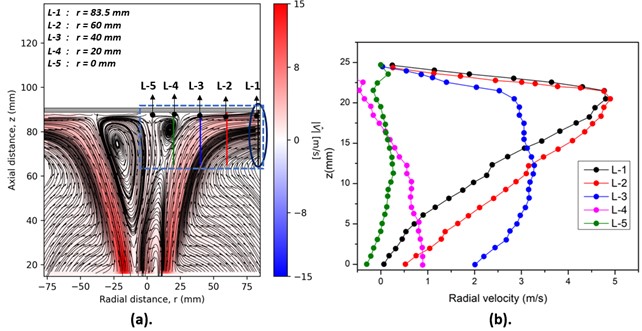}
\caption{Radial velocity after impingement along the plate radial direction from PIV experiment (a). Contour streamline plots with lines L-1, L-2, L-3, L-4, and L-5 along the impingement plate for probing (with axial distance of 25mm) the radial velocity profiles (b). Radial velocity profiles at different lines for H/D = 3 at Re=23000.}
\label{fig:Fig14}
\end{figure}

\begin{figure}[h!]
\centering
\includegraphics[scale = 0.9]{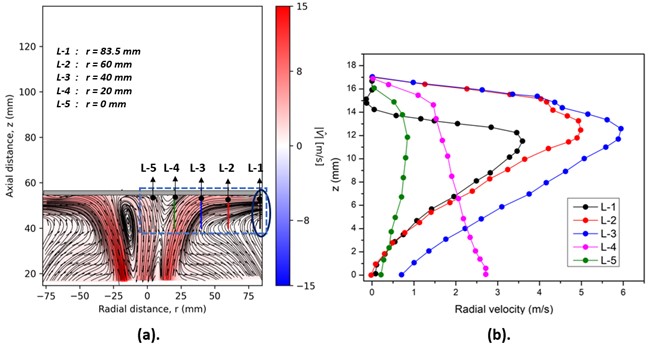}
\caption{The radial component of velocity at lesser impingement distance H/D = 2 at Re=23000.}
\label{fig:Fig15}
\end{figure}

\subsubsection{Near wall azimuthal velocity profiles upon impingement}
Apart from the radial velocity components at the near wall upon impingement, there is also considerable azimuthal velocity magnitude, which may add to considerable impact on the transport properties (heat) in the swirl jet impingement, which is absent in conventional non-swirl (or) round jets. Here, the azimuthal and radial velocities are computed near the wall at two radial locations near wall jets L-1 and L-2 for two jet-plate distances H/D = 3 and 2 at Re=23000.

\begin{figure}[h!]
\centering
\includegraphics[scale = 0.85]{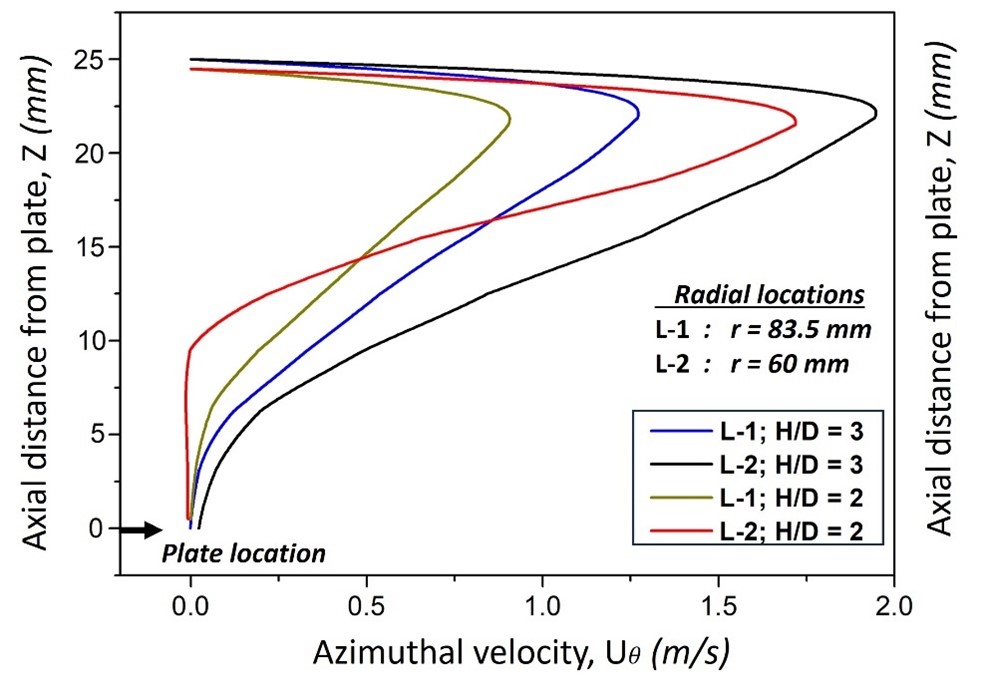}
\caption{Computed near wall azimuthal velocity profiles at Re=23000 and the effect of H/D on its magnitude is depicted at H/D = 2 and 3. The data are plotted for 2 radial locations L-1 and L-2 which corresponds to r/D = 2.78, and 2.}
\label{fig:Fig16}
\end{figure}

\newpage

\subsection{Influence of jet-plate distance(H/D) on the re-circulation topology}
The jet-plate distance is an important parameter to be considered in the jet impingement study. Literature suggests that as this distance H/D, the effect of swirl decreases. 

\begin{figure}[h!]
\centering
\includegraphics[scale = 1]{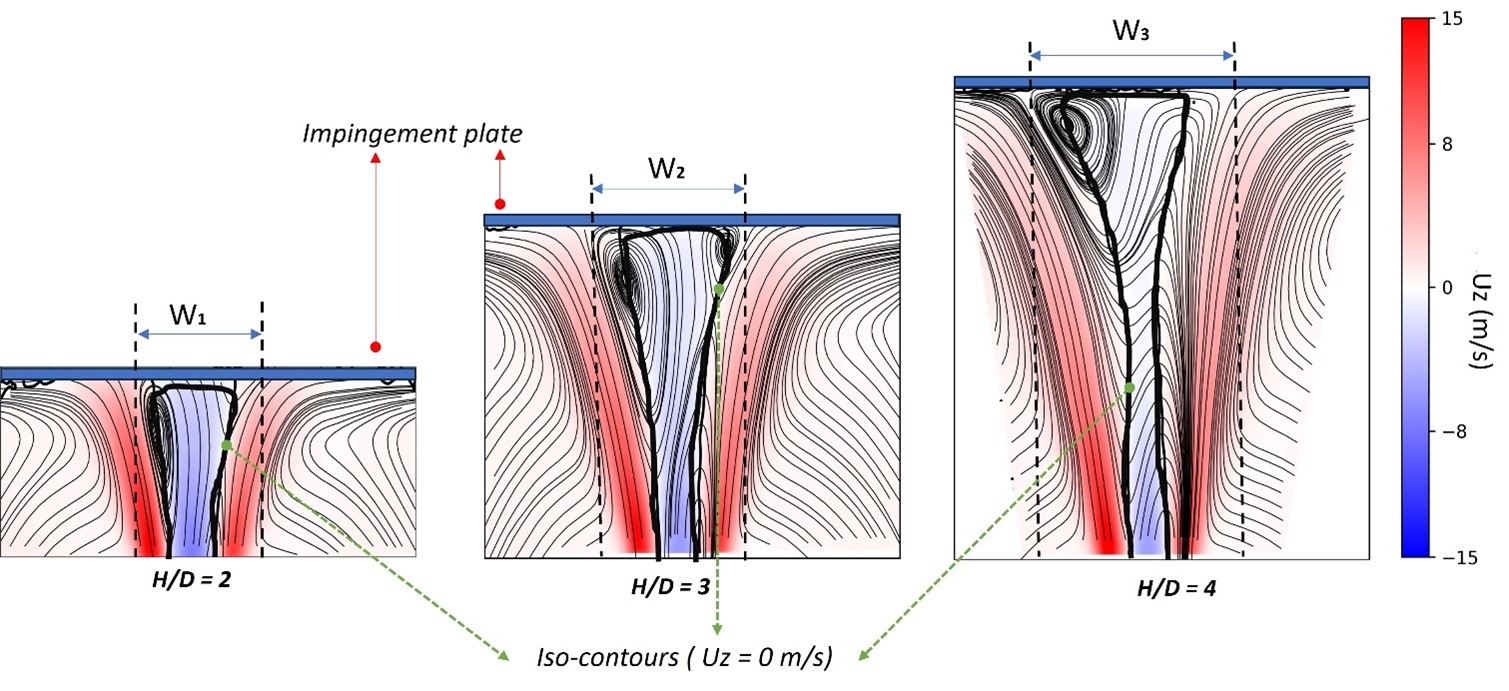}
\caption{Width of the re-circulation zones at different impingement distances H/D = 2,3, and 4 corresponding to the Reynolds no. (Re = 23000) from the PIV experiment. The iso-contour (Uz = 0) indicating zero axial velocity has been marked for each case.}
\label{fig:Fig17}
\end{figure}

The above figure is an illustration of the effect of impingement distance on the flow structure of the swirl jet impinging on the flat plate. For all the cases, the recirculating vortex extends up to the impingement distance or the location of the plate. With the increase in the impingement distance (H/D), the width of the recirculating vortex increases; in other words, the topology of the vortex increases, which can also be quantified in terms of L/D. The iso-contour for zero axial velocity (Uz = 0) which is analogous to a shear layer demarcating the regions of positive velocity from the negative velocity region (i.e. recirculation zone) are also shown. 

\subsection{Effect of jet-plate distance(H/D) on the velocity}
The effect of H/D on the velocity components at far and near field to the plate is understood by plotting the velocity components at 2 locations (a). centre of the FOV (z = 45mm or H/D = 1.5) (b). Near the impingement plate (z = -5 mm from the plate). Near the plate, the velocity components vary much for both axial and radial velocity components.
\begin{figure}[h!]
\centering
\includegraphics[scale = 0.85]{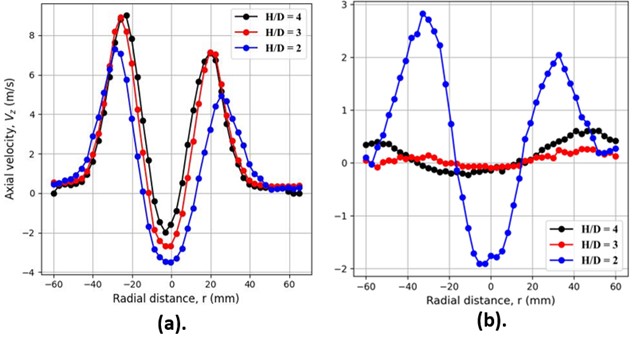}
\caption{Effect of H/D on the axial velocity components for each case of H/D = 4, 3, and 2 at Re = 23000 (a). At a far field from the plate (at z = 45mm from the jet exit) (b). At a near field to the plate (5 mm below the impingement plate).}
\label{fig:Fig18}
\end{figure}

\begin{figure}[h!]
\centering
\includegraphics[scale = 0.85]{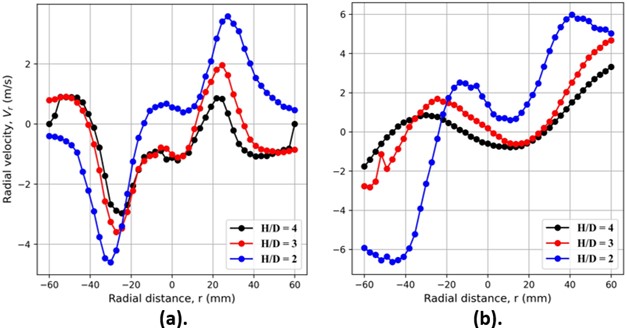}
\caption{Effect of H/D on the radial velocity components for each case of H/D = 4, 3, and 2 at Re = 23000 (a). At a far field from the plate (at z = 45mm from the jet exit) (b). At a near field to the plate (5 mm below the impingement plate).}
\label{fig:Fig19}
\end{figure}


 \begin{figure}[h!]
\centering
\includegraphics[scale = 0.9]{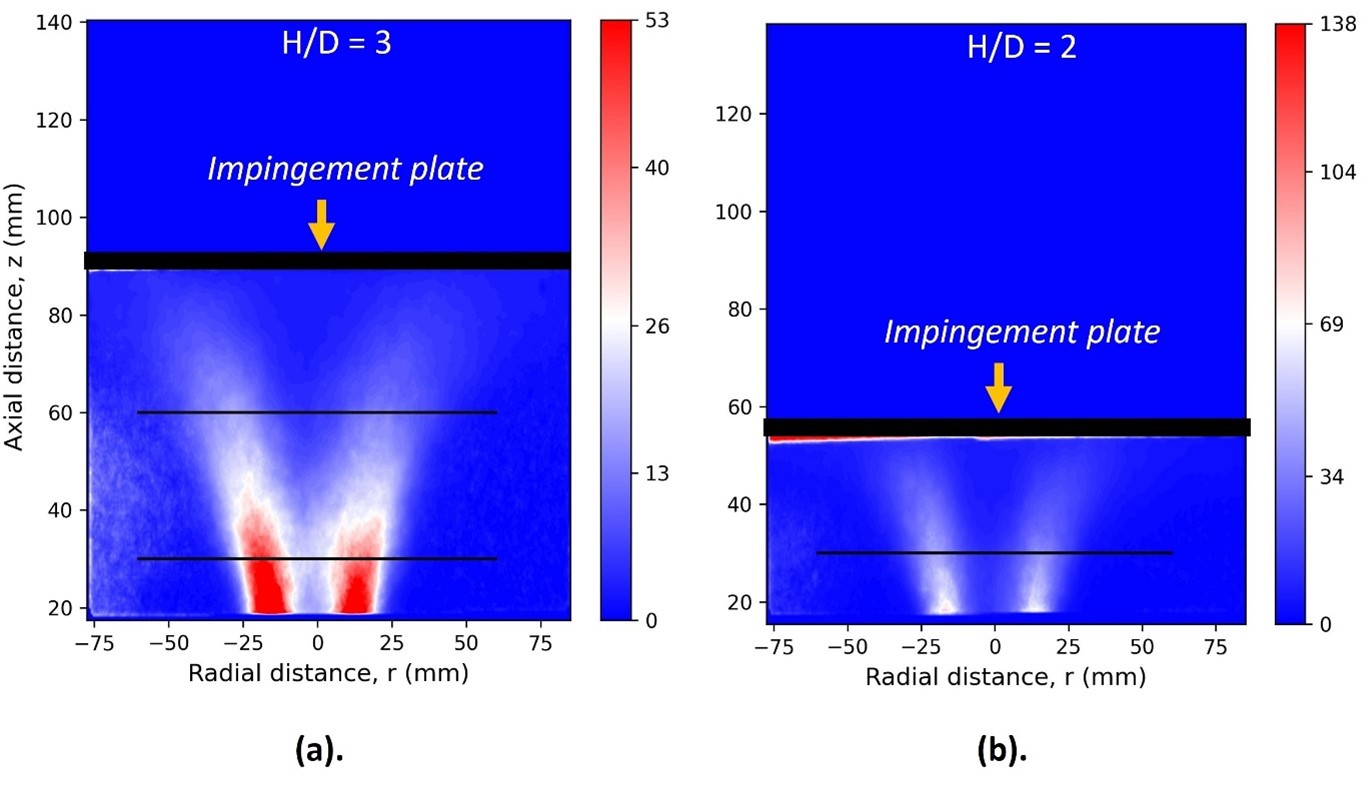}
\includegraphics[scale = 0.85]{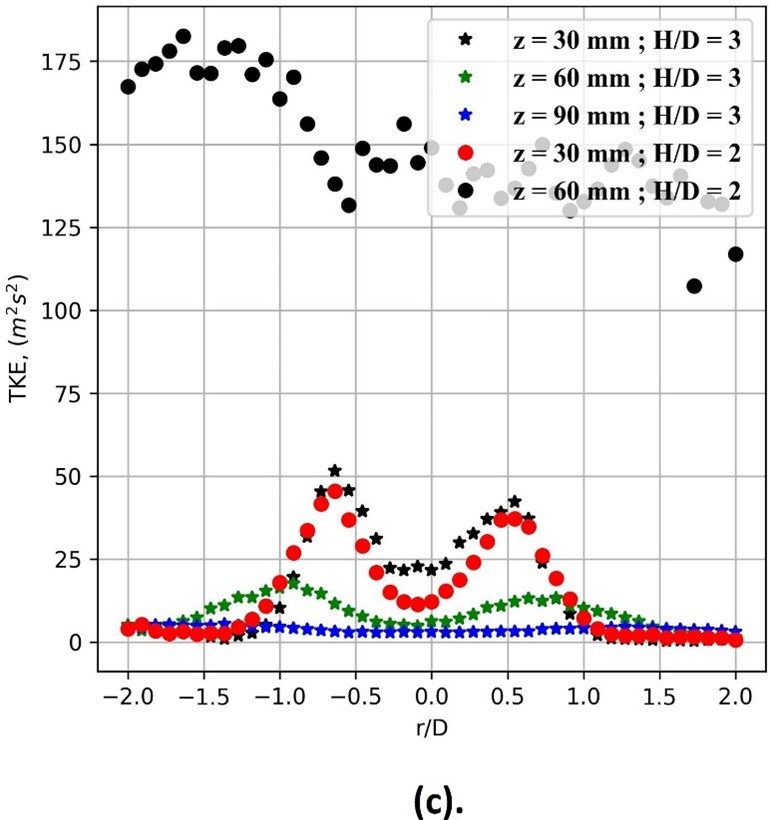}
\caption{Effect of jet-plate distance (H/D) on the turbulence kinetic energy (TKE) is depicted from PIV experiments (a). Contour plots for TKE at impingement distance H/D = 3 (b). At impingement distance H/D = 2. (c). TKE (line data) plotted at various axial distances (z = 30,60, and 90mm) for the cases H/D = 3 and 2.}
\label{fig:Fig20}
\end{figure}

\subsection{Impingement turbulence statistics}
Swirl jet impingement is characterized by complex turbulence dynamics that influence mixing, momentum transfer, and heat transfer rates.

The three key turbulence parameters, namely the Turbulence Kinetic Energy (TKE), RMS fluctuating velocity, and Reynolds Stress Tensor, quantify different aspects of turbulent motion and together provide a comprehensive understanding of the flow physics. This section discusses the above-mentioned turbulence statistics for the impinging swirl jet. For jet impingement studies, the inlet Reynolds number (Re), and jet-plate distance (H/D) are the obvious parameters that influence these turbulence quantities. As the effect of Re is a trivial case, we study only the effect of H/D on the impingement turbulence statistics.

 \subsubsection{Significance of TKE, RMS fluctuating velocity, and Reynolds Stress Tensor in swirl jet impingement with varying jet-plate distance (H/D)}
 
The effect of H/D on the average turbulence characteristics, such as Turbulence kinetic energy, Reynolds stress tensor, and fluctuation root mean square (RMS) velocity components, is discussed at two jet-plate distances, H/D = 3 and 2. The idea is to show that at a lower jet-plate distance, the TKE and RSS are more pronounced, which could be the reason for an enhanced heat transfer for swirl jet at lower impingement distances $(H/D < 2)$.\\
\\
\textbf{(a). Turbulence Kinetic Energy (TKE):}

TKE represents the total energy associated with turbulent velocity fluctuations and is a key parameter in assessing turbulence intensity. At lower impingement distance $(H/D \leq 2)$, the swirl jet undergoes premature stagnation, resulting in weaker TKE in the shear layer due to weaker vortex breakdown. However, the near wall turbulence is higher due to strong flow deceleration as stated by round jet impingement literature (Cooper et al, 1993; Hall and Ewing, 2006). In Fig.20. (c). For the H/D = 2 or H = 60mm case, the TKE near the impingement plate is about an order of magnitude higher compared to other locations. However, for the H/D = 3 case, the TKE near the impingement plate i.e. near wall turbulence is much weaker than the rest of the locations shown in the Fig.20. (c) which can be due to the decay of swirl and turbulence at this height in par with the literature [W.G.Ross]. This ascertains that for the swirl jets, the turbulence energy at lower jet-plate distances is pronounced, which also supports our reasoning for enhanced heat transfer.
\\
\\
\textbf{(b). Reynold's stress tensor:}

The momentum transport between different layers of fluid due to turbulent fluctuations in turbulent flow is quantified by the Reynolds stress tensor, also known as Reynolds shear stress (RSS). 

\begin{figure}[h!]
\centering
\includegraphics[scale = 0.9]{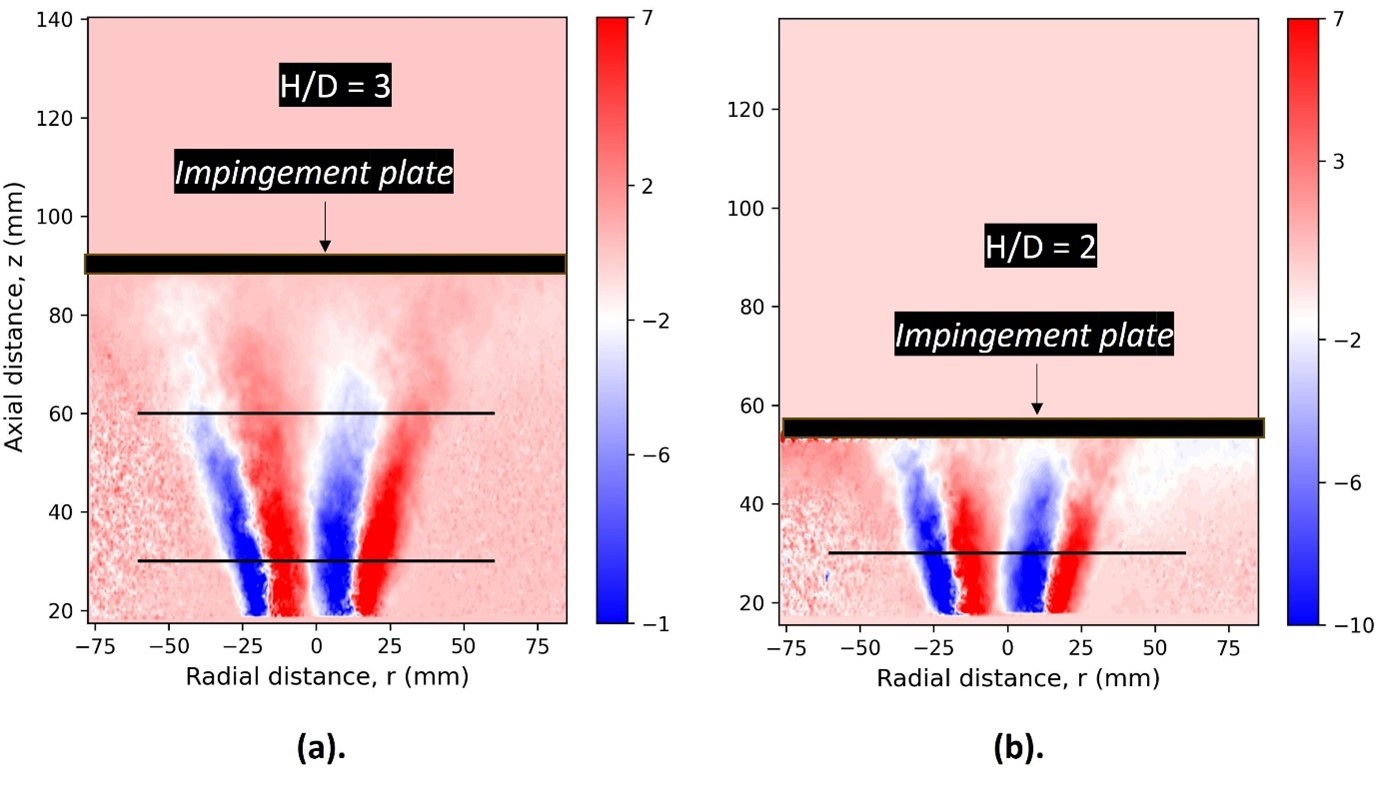}
\includegraphics[scale = 0.85]{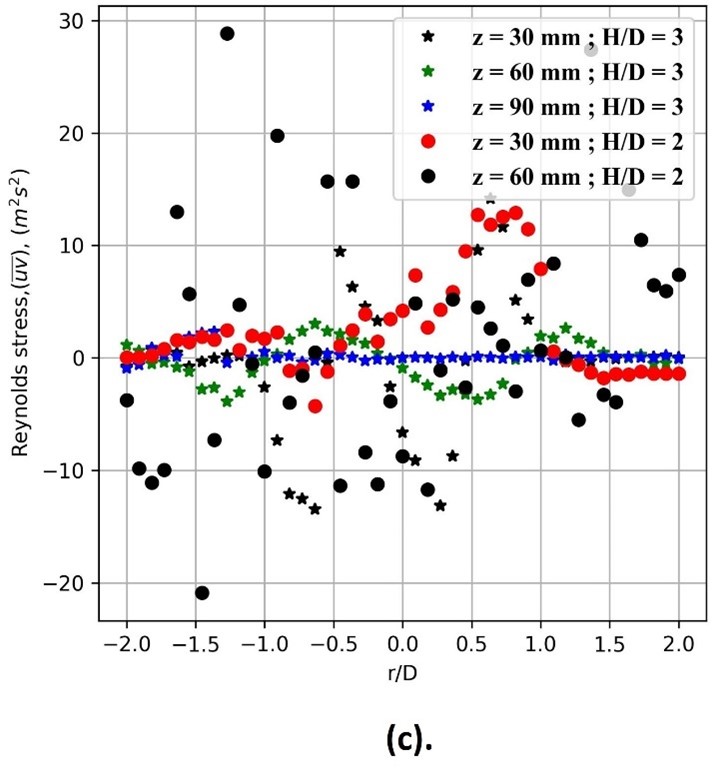}
\caption{Reynold’s shear stress RSS (fluctuating mean velocity components) for the swirling jet impinging the plate at H/D = 2 or H = 60 mm corresponding to Re = 23000. (a). Contour plots for RSS at impingement distance H/D = 3 (b). At impingement distance H/D = 2. (c). RSS (line data) plotted at various axial distances (z = 30,60, and 90mm) for the cases H/D = 3 and 2.}
\label{fig:Fig21}
\end{figure}

For the impinging swirl jets, especially at small jet-plate distances $(H/D\leq2)$, strong velocity gradients increase shear stresses $\overline{u'v'}$ near the wall, which in turn enhances localised turbulence production. Also, the swirl-induced secondary vortices generate higher azimuthal stresses $\overline{w'w'}$ [Cooper et al, 1993]. However, this cannot be shown for the PIV experimental case, which had limitations.

The above figure, Fig.21 (c), shows that the Reynolds shear stress is concentrated more near the impingement plate for the case H/D = 2, as the momentum transfer is greater at the event of impingement near the plate. For H/D = 3 case, the RSS are evident up to an axial distance 2D i.e., z = 60mm beyond which the swirl almost decays, that again justifies an enhanced turbulence at a lower jet-plate distances for swirl jets. \\
\\

\textbf{(c).Fluctuating RMS velocity:}

 Fluctuating RMS describes how much the turbulent velocity deviates from the mean flow at a given moment, with a higher RMS indicating greater turbulence intensity. Both the axial and radial RMS, i.e., URMS and VRMS, are reported to be higher near the plate (Fig.23) for the H/D = 2 case, with $ U_{RMS} > V_{RMS}$. The fluctuating RMS components had a distinct profile like the velocity components. Here, we can also conclude that at lower jet-plate distances, the RMS velocities are higher, which may lead to a fluctuating or unsteady type of heat transfer in practical applications.
\begin{figure}[h!]
\centering
\includegraphics[scale = 0.9]{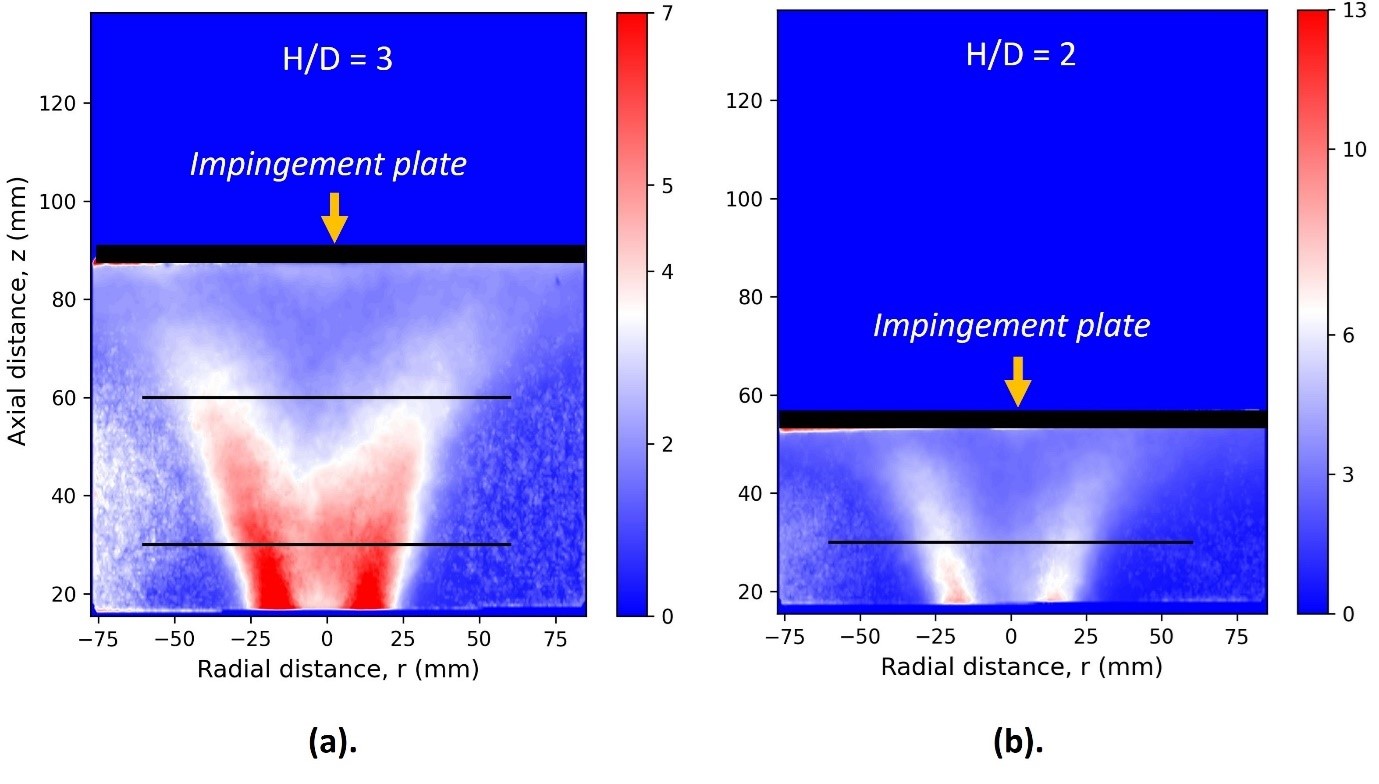}
\caption{Effect of H/D on the RMS velocity for the impinging jet at Re=23000 (a). Contour plot showing the magnitude of fluctuating velocity at H/D = 2 (b). Contour plot at H/D = 3.}
\label{fig:Fig22}
\end{figure}

\begin{figure}[h!]
\centering
\includegraphics[scale = 0.85]{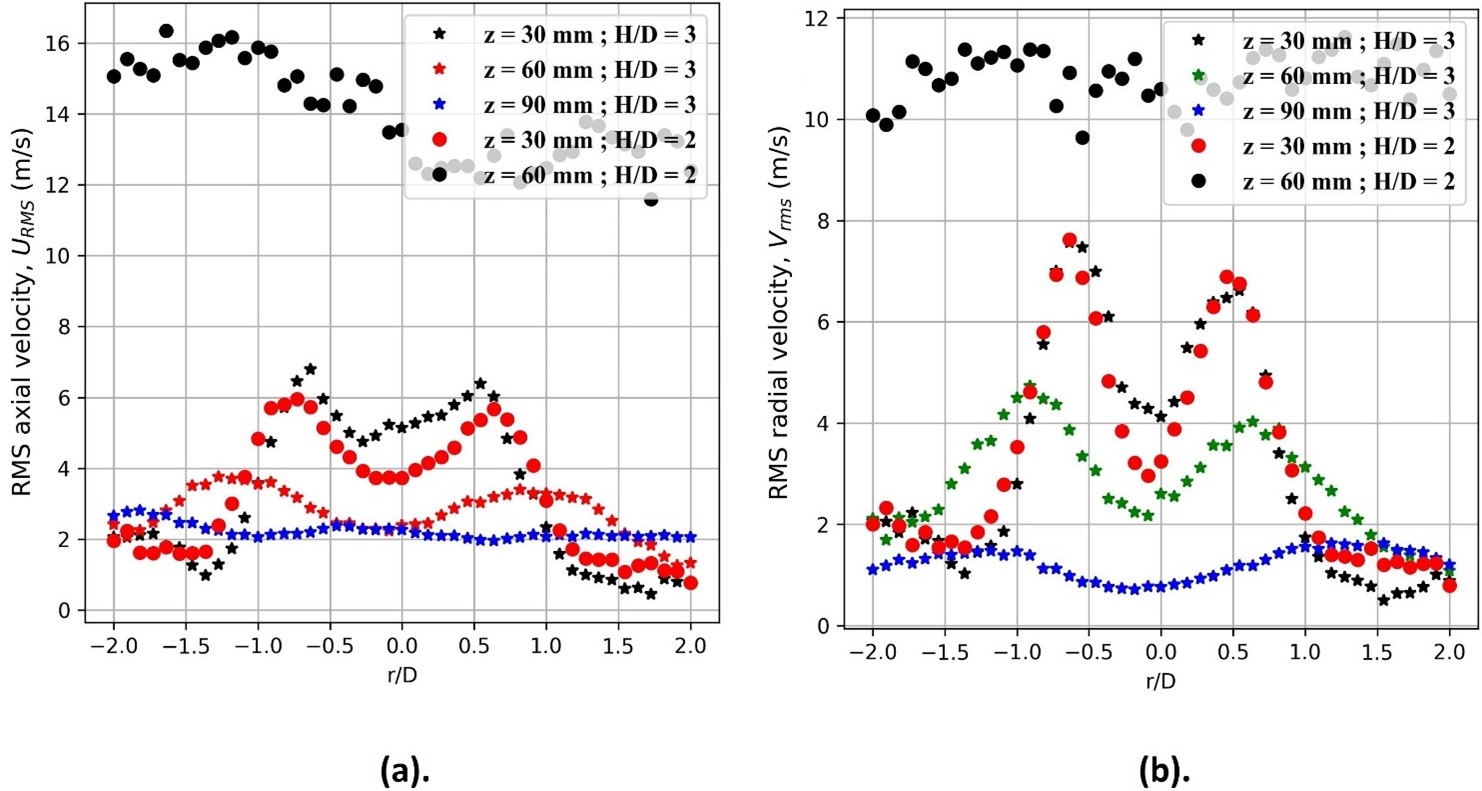}
\caption{Fluctuating RMS velocity components line data at different axial velocities at Re=23000; H/D = 2, and 3 (a) Axial components (URMS) (b). Radial components (VRMS) at different axial locations z = 30, 60, and 90 mm.}
\label{fig:Fig23}
\end{figure}

\subsection{Proper Orthogonal Decomposition}
Proper Orthogonal Decomposition (POD) is widely used to analyse turbulent flows by extracting dominant coherent structures from experimental or numerical datasets.

\begin{figure}[h!]
\centering
\includegraphics[scale = 0.90]{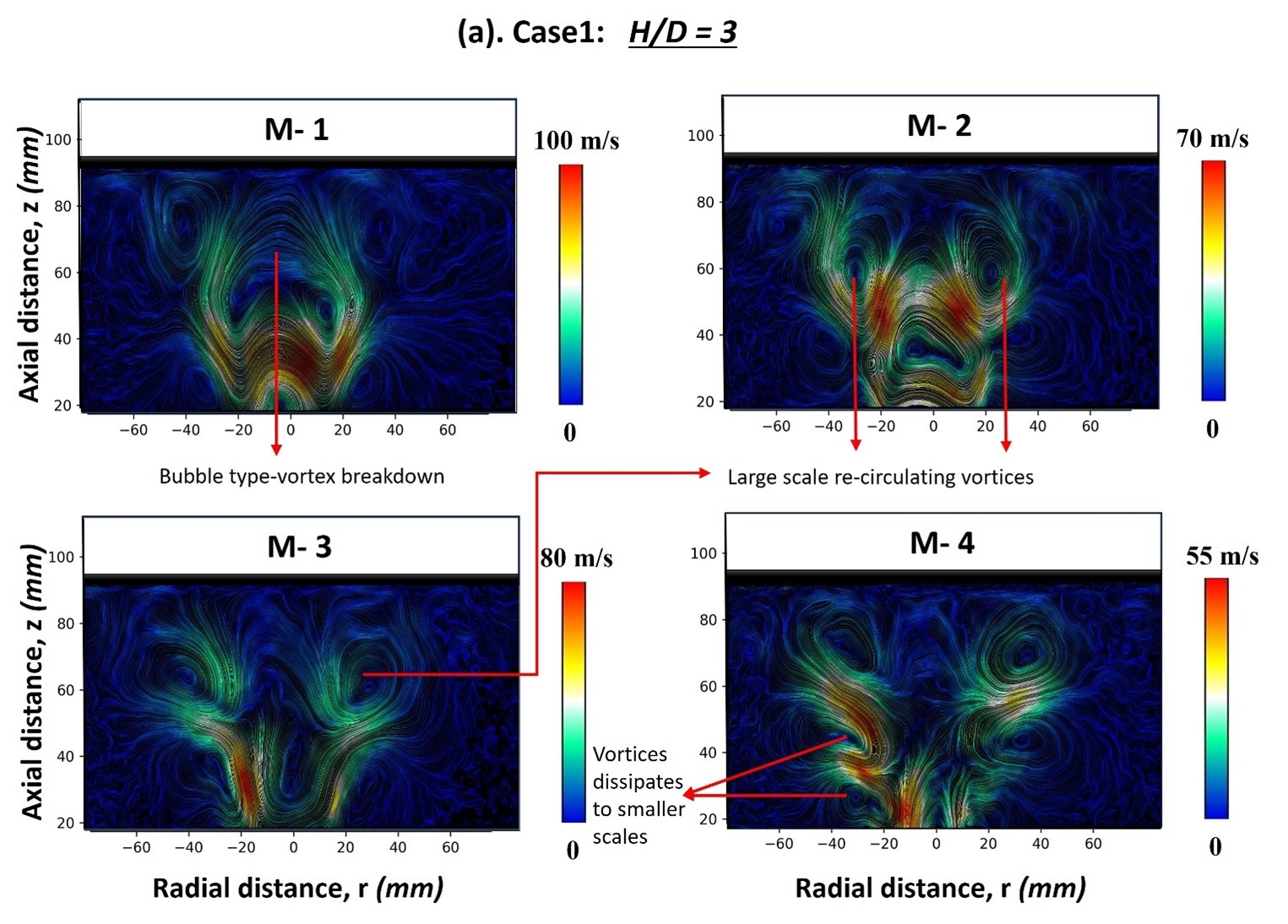}
\includegraphics[scale = 0.95]{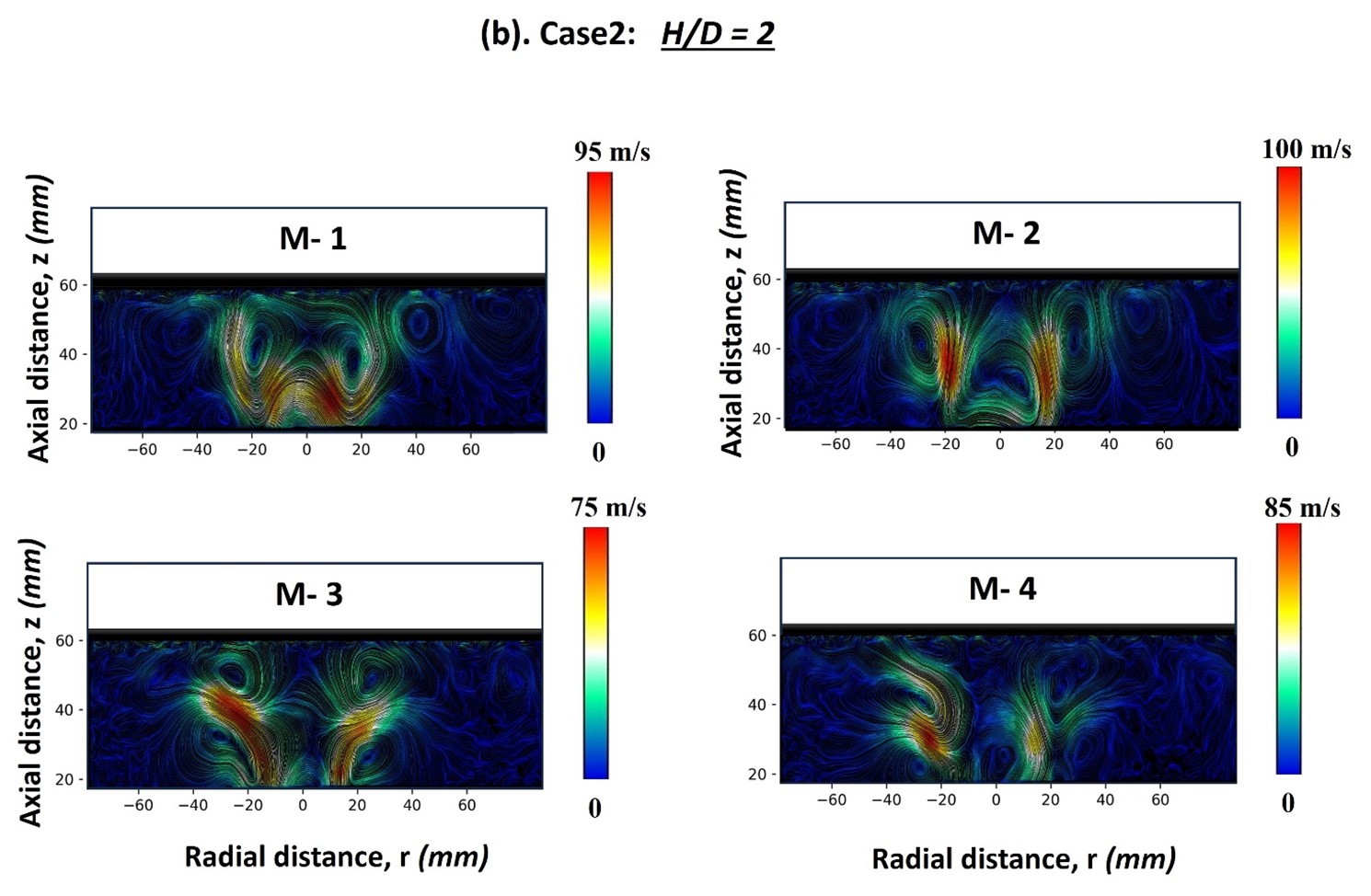}
\caption{POD modes (first 4 dominant modes) for the impingement cases at (a). H/D = 3 (b). H/D = 2.}
\label{fig:Fig24}
\end{figure}

\begin{figure}[h!]
\centering
\includegraphics[scale = 0.90]{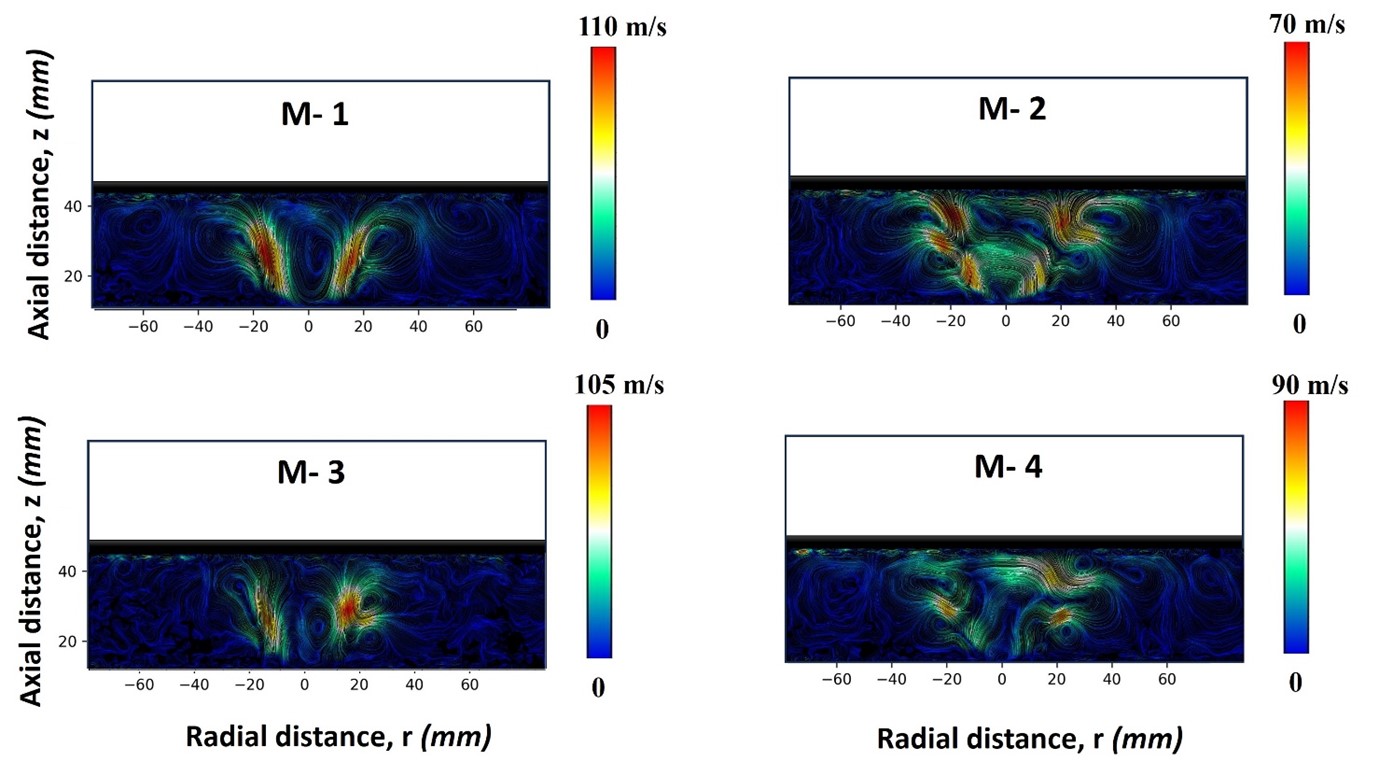}
\caption{POD modes (first 4 dominant modes) for the impingement cases at H/D = 1.5 }
\label{fig:Fig24.c}
\end{figure}
 In swirl jet impingement, POD is particularly useful for identifying large-scale vortices, analysing vortex breakdown, and quantifying turbulence energy distribution.
This section discusses the evolution of the spatial modes of the swirl jet impinging on the flat plate at Re=23000 for the jet-plate distances (H/D = 3, 2, and 1.5) using POD. 
From the instantaneous velocity fields dataset obtained from PIV, the proper orthogonal decomposition (POD) decomposes the instantaneous velocity field $ u(x,t) = \Sigma_{i=1} ^N a_{i}(t) \phi_{i} (x)$, \\
where $\phi_{i}(x)$ is the orthogonal spatial modes or eigen modes which represent coherent structures in the flow, $a_{i} (t)$ are the temporal coefficients which describe the time-dependent behaviours of each mode. N is the total number of modes (equal to the number of snapshots).
The POD calculation involves the computation of the covariance matrix of the velocity data from the column matrix containing all the velocity data. Eigenvalues and eigenvectors are extracted from the covariance matrix, where the eigenvalues represent the energy contribution of each mode. 

\begin{figure}[h!]
\centering
\includegraphics[scale = 1]{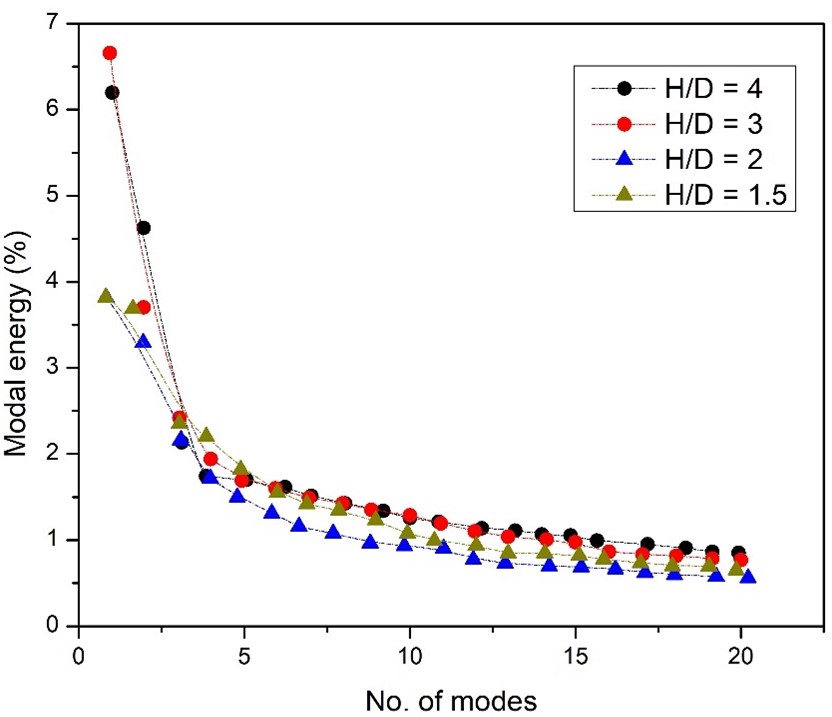}
\caption{Percentage of modal energy at different jet-plate distances (H/D).}
\label{fig:Fig25}
\end{figure}

Here, the POD analysis is performed, showing the coherent structures for the first 4 modes for three cases of H/D = 3,2, and 1.5. Generally, the most energetic modes (the first few dominant modes) represent the large-scale coherent structures in the flow. In Fig.24 (a), which represents the first 4 modes for the case H/D = 3, the dominant first mode indicates the bubble-type vortex breakdown, which is a common type of vortex breakdown mechanism in swirl jets. The next two modes, i.e., modes M-2 and M-3, show the large-scale vortex structures and their evolution. The mode M-4 gives some hint for the dissipation of larger vortices into smaller vortices. 
 Surprisingly, for the other jet-plate distances H/D = 2, and 1.5 which are at lower heights, we can observe a self-similar coherent structure for all the modes M-1,2,3, and 4 with a difference only in the sizes of the vortex structures (refer Fig.24(b), (c)). This makes us reason out that adjusting the impingement plate to lower distances compresses and deforms the flow structures to scaled-down coherent vortex structures, which is quite an interesting observation.
Fig.25 shows the percentage modal energy plotted at different modes (up to 20 modes) for the different cases of jet-plate distances H/D. 

\section{Conclusion}
In the present work, both experimental 2-D PIV and 3-D computational studies are performed to investigate the swirl jet impingement flow field and topology. The 2-D PIV work had experimental flow visualisation investigations at both the front $(r-z)$ and top $(r-\theta)$ planes. The 3-D numerical simulations are performed in a place where experimental techniques had practical limitations. A comparative study of the flow topology of the swirling jet at impinging and free states (in the appendix) is presented. The influence of jet–plate distance (H/D) and Reynolds number on the swirl jet flow topology for these jets is discussed. The following are the important conclusions,\\
\begin{itemize}
    \item The near-wall azimuthal velocity components, apart from the radial velocity components, which are important for the wall jet convective transport, are elucidated.
    \item The effect of jet-plate distance on the impinging jet topology is shown.
    \item A 3-D flow structure from both experimental and computational results is presented. In particular, the vortex breakdown of the swirl jet and the distortion of shear layers are represented from the computational results, which closely match the PIV experimental studies.
    \item The turbulent kinetic energy is found to be an order higher for the swirl jet impinging at the low jet-plate distance $(H/D\leq2)$ concentrated at the near-wall of the impingement plate.
    \item Proper orthogonal decomposition (POD) on the dominant spatial modes is presented for the impinging jet cases at H/D = 3,2, and 1.5 and their modal energies are plotted to show that the maximum energies in the impingement jet flow field are concentrated at lower jet-plate distances. 
\end{itemize}

\newpage

\textbf{Appendix A: Flow topology of free swirl jet from the 2D-PIV experiment}\\
Here, the flow topology of the free swirl jet (non-impingement) from the PIV experiment and 3D computational simulations is shown and compared. The key flow features of a typical swirl jet are characterized by a vortex-induced re-circulation zone, followed by the internal stagnation points (leading and trailing) and the shear layers. In this case, a bubble-type vortex breakdown is formed, which leads to a central re-circulation bubble (CRZ). The above-mentioned key flow features are indicated in the Figure, which depicts the velocity contour plot superimposed with the streamlines along the mid-longitudinal or front plane from the 2D-PIV experiment carried out at Re = 16000.
\begin{figure}[h!]
\centering
\includegraphics[scale = 0.85]{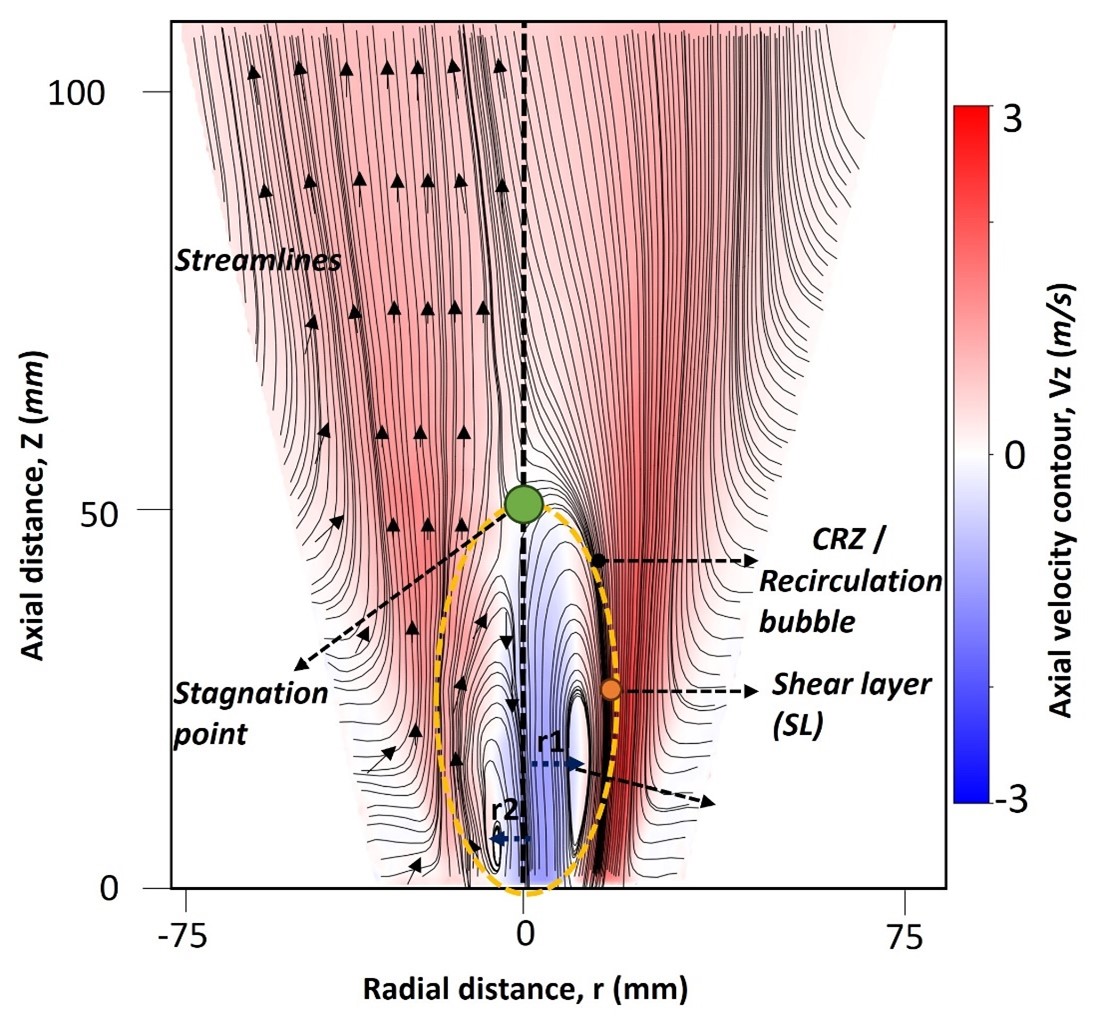}
\caption{Streamline and velocity contour (axial component) plot along the front plane (r-z) for a 450  swirl jet corresponding to the Reynolds no. (Re = 16000) from the PIV experiment corresponding to free jet (non-impingement).}
\label{fig:Fig26}
\end{figure}

To understand the flow topology at the top $(r-\theta)$ planes, 2D-PIV experiments are performed at two $(r-\theta)$ planes at vertical distances z = 45mm and 90mm. Owing to the optical constraints associated with the CCD camera, experiments can’t be performed at lower distances below z = 45mm, however literature suggests that prominent swirl strength are concentrated at a low axial distances (between z = 1 – 2 D, where D is the jet diameter) and beyond this range the swirl decays. Figure 28.(a) \& (b) depict the averaged streamlines and velocity contour at the top planes z = 45mm and 90mm, respectively. In the z = 45mm case, there are evident shear layers that may correspond to inner and outer shear layers (ISL \& OSL) respectively. We can confirm a predominant swirling motion within this shear layer, as the strength of the swirl is strong for this case. At a higher distance $  r-\theta$ plane, at z = 90mm, there is a shifting of the shear layer from the centre to the circumference or periphery, indicating the decay of swirl with a decreased swirl strength.

\begin{figure}[h!]
\centering
\includegraphics[scale = 0.85]{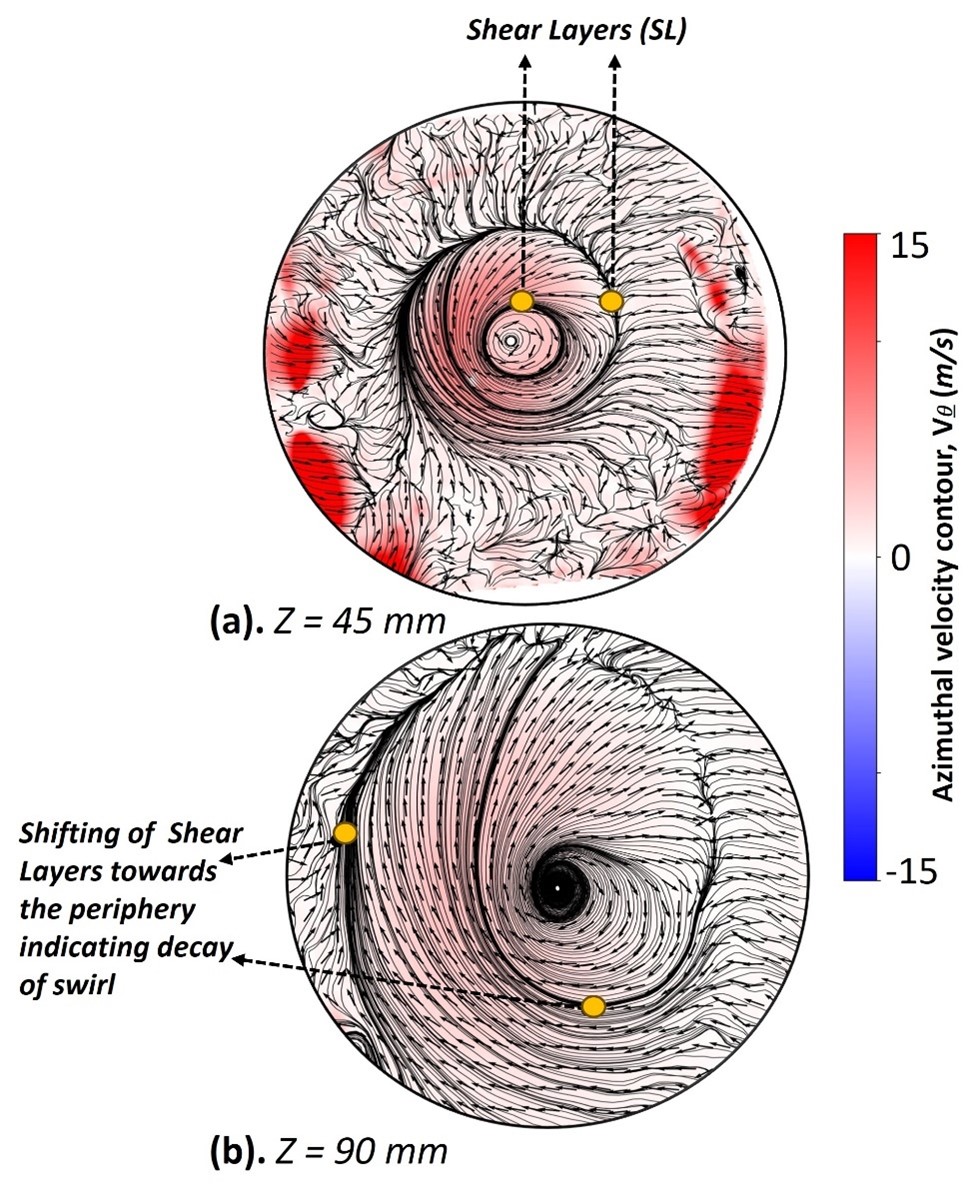}
\caption{Streamline and average velocity contour plot corresponding to Re = 23000 at the top $(r- \theta)$ planes (a). z = 45mm, and (b). z = 90mm, indicating the decay characteristics of swirl flow along the axial direction.}
\label{fig:Fig27}
\end{figure}

\newpage

\textbf{Appendix B: Flow topology of free swirl jet from the computational studies:}\\
As we seek the aid of 3D-computational simulations where experiments have practical limitations in the impinging cases, it is important to validate the current flow-field computational results with the PIV experiments. The validation plots for the velocity components for the free and impingement cases have already been shown in Section 3.1. Here, we show the computed flow topology for the free jet, very similar to the experimental flow structures shown in Appendix A.
Figure 29. shows the computed streamlines for the Re = 16600 at $r-\theta$ planes z = 45mm and 90mm, respectively. In par with the experimental results depicted in Fig.27, the computation also confirms the ISL and OSL and shifting of the shear layers from the centre to the periphery, as can be verified in Figs . 29 and 30. 
Also, Fig.30 shows the streamlines from the front plane, which depicts the bubble vortex breakdown-induced re-circulation zone, shear layer, and internal stagnation point.
\begin{figure}[h!]
\centering
\includegraphics[scale = 0.85]{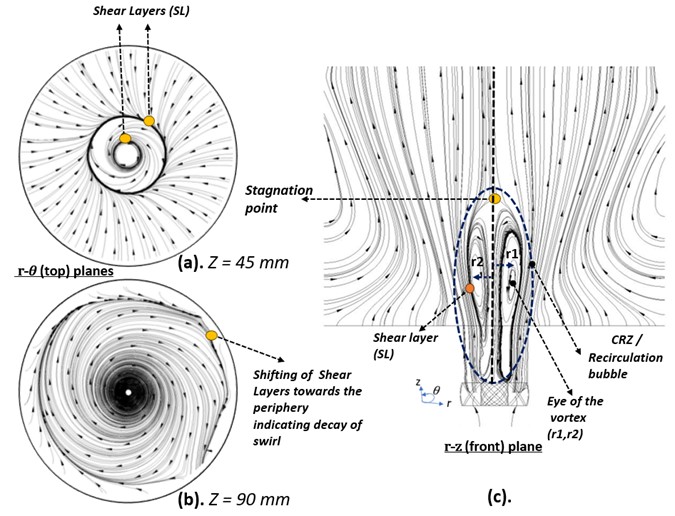}
\caption{Computed streamlines for the top $(r-\theta)$ plane similar to the above experimental PIV results at two different axial locations (a). z = 45 mm, and (b). z = 90 mm, and at (c). Front r-z plane (mid-longitudinal).}
\label{fig:Fig28}
\end{figure}

The below figure.B.2. shows the 3D streamlines which depict the vortex breakdown of the bubble type. The axisymmetric type of vortex breakdown (bubble type), which is the most common type of in swirl flows [Leibovich (1978)], can be confirmed from the Figs.30 and 29(c)

\begin{figure}[h!]
\centering
\includegraphics[scale = 0.95]{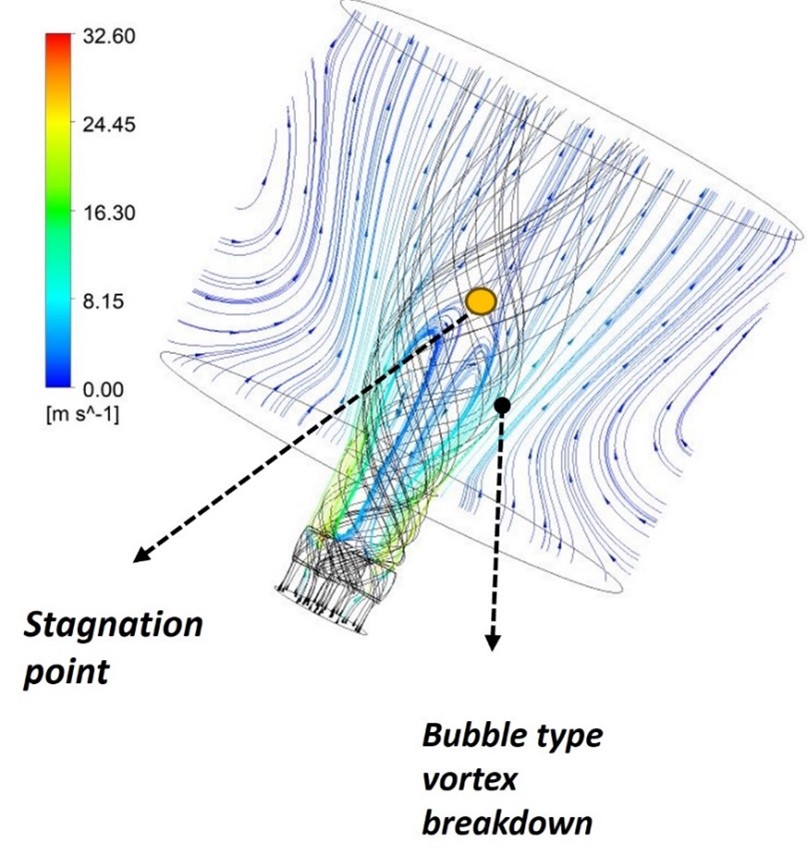}
\caption{Computed 3D streamlines superimposed over front r-z (mid-longitudinal) plane streamlines depicting the bubble-type vortex breakdown, recirculating streamlines, and stagnation point.}
\label{fig:Fig29}
\end{figure}

\newpage

\textbf{Appendix C: Measure of asymmetry from the Velocity decay characteristics:}\\
The decay of axial velocity profiles along the centreline and along the left and right lines, each at 15mm from either side of the centreline, is shown for the impinging and non-impinging (free jet) cases. The deviation between the velocity profiles along the left and right lines is used as a measure of asymmetry in the flow here.
\begin{figure}[h!]
\centering
\includegraphics[scale = 0.95]{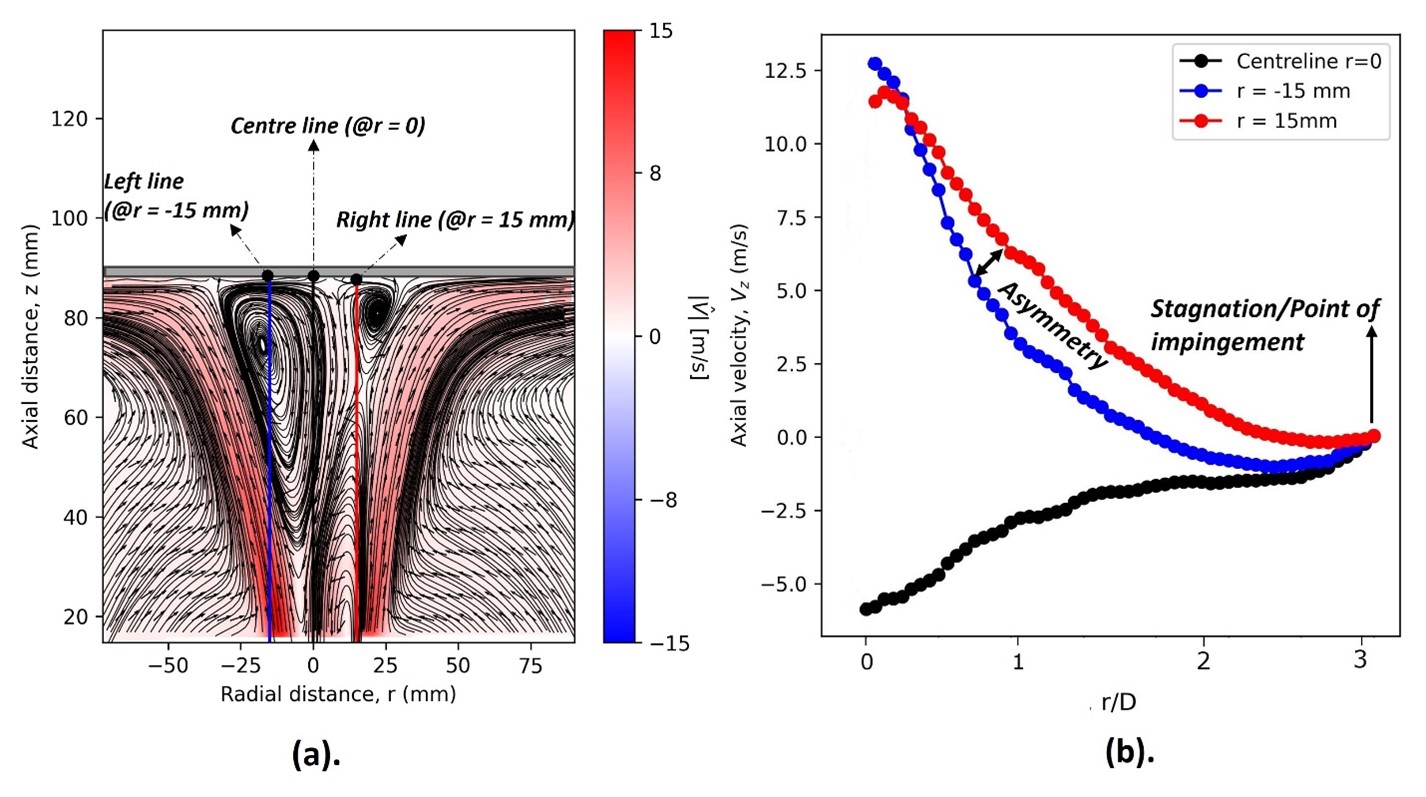}
\includegraphics[scale = 0.95]{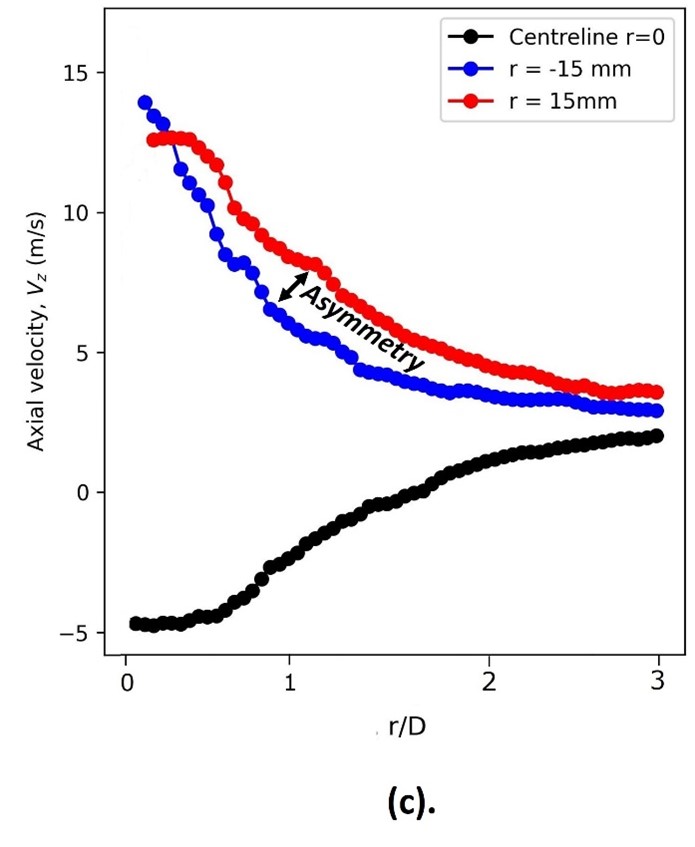}
\caption{Axial velocity (centreline) decay characteristics for an impinging at H/D = 3 and freely swirling jet (Re=23000) (a). Average velocity contour and streamlines with indication of data probing lines (left, centre and right lines) (b). Axial velocities for the impinging jet case, and (c). Axial velocities for the non-impinging (free jet case) along the centreline (r = 0), left and right lines (r = -15 mm, r = 15mm) and the deviation in velocity characteristics between the left and right line indirectly indicate the measure of asymmetry.}
\label{fig:Fig30}
\end{figure}

\newpage

\textbf{Acknowledgement:}\\
The authors acknowledge Darshan Rothod for assistance with Experiments and the GTCRL laboratory facility of IISc Bangalore. Special thanks to Dr.Pratikash Panda for his valuable suggestions. \\
\\

\textbf{References:}
\\

[1] Martin H, Heat and Mass Transfer between Impinging Gas Jets and Solid Surfaces, Advances in Heat Transfer,13 (1977), 1-60.\\

[2] R.J. Goldstein, A.I. Behbahani, Impingement of a circular jet with and without cross flow, Int. J. Heat Mass Transf. 25(1982), 1377-1382.\\

[3] Donaldson C. duP, Snedeker RS. A study of free jet impingement. Part 1. Mean properties of free and impinging jets. Journal of Fluid Mechanics. 1971;45(2):281-319. \\

[4] Gardon, R. \& Akfirat, J. C., The role of turbulence in determine the heat transfer characteristics of impinging jets, Int. J. Heat Mass Transf. 8 (1965), 1261-1272.\\

[5] Cooper, D., Jackson, D. C., Launder B. E. and Liao G. X. 1993 Impinging jet studies for turbulence model assessment – I. Flow-field experiments. Int. J. Heat mass Transfer 36 (10), 2675-2684. \\

[6] M. Hadziabdic, K. Hanjalic, Vortical structures and heat transfer in a round impinging jet, J. Fluid Mech. 596 (2008) 221–260.\\

[7] Nicholas Syred, A review of oscillation mechanisms and the role of the precessing vortex core (PVC) in swirl combustion systems, Progress in Energy and Combustion Science, Volume 32, Issue 2, 2006, Pages 93-161. \\

[8] Fairweather, M. \& Hargrave, G. K. 2002 Experimental investigation of an axisymmetric, impinging turbulent jet. 1. Velocity field. Experiments in Fluids 33, 464-471.\\

[9] F. Gallaire, J.-M. Chomaz; The role of boundary conditions in a simple model of incipient vortex breakdown. Physics of Fluids 1 February 2004; 16 (2): 274–286.\\

[10] D Lytle, B.W Webb (1994), Air jet impingement heat transfer at low nozzle-plate spacings, Int J Heat and Mass Transfer, 37.12, 1687-1697.\\

[11] Hassan, Syed Harris, Guo, Tianqi, and Vlachos, Pavlos P. (2019). “Flow field evolution and entrainment in a free surface plunging jet”. Phys. Rev. Fluids 4 (10),104603.\\

[12] M. A. Herrada, C. Del Pino, J. Ortega-Casanova; Confined swirling jet impingement on a flat plate at moderate Reynolds numbers. Physics of Fluids 1 January 2009; 21 (1): 013601.\\

[13] Rose, W. G., A Swirling Round Turbulent Jet: 1—Mean-Flow Measurements. ASME. J. Appl. Mech. December 1962; 29(4): 615–625. \\

[14] Ekkad, S. V., and Singh, P. (April 21, 2021). "A Modern Review on Jet Impingement Heat Transfer Methods." ASME. J. Heat Transfer. June 2021;143(6):064001.\\

[15] Gómez-Ramírez, David et al. “Isothermal coherent structures and turbulent flow produced by a gas turbine combustor lean pre-mixed swirl fuel nozzle.” Experimental Thermal and Fluid Science 81 (2017): 187-201.\\

[16] Sergey V. Alekseenko, Artur V. Bilsky, Vladimir M. Dulin, Dmitriy M. Markovich, Experimental study of an impinging jet with different swirl rates, Int J Heat and Fluid Flow, 28(6), 1340-1359. \\

[17] Liang Xu, Tao Yang, Yanhua Sun, Lei Xi, Jianmin Gao, Yunlong Li, Jibao Li, Flow and heat transfer characteristics of a swirling impinging jet issuing from a threaded nozzle, Case Studies in Thermal Engineering, Volume 25, 2021, 100970, https://doi.org/10.1016/j.csite.2021.100970. \\

[18] L. F. G. Geers, M. J. Tummers, K. Hanjalić; Particle imaging velocimetry-based identification of coherent structures in normally impinging multiple jets. Physics of Fluids 2005; 17 (5): 055105. \\

[19] L. F. G. Geers, M. J. Tummers, K. Hanjalić Wall imprint of turbulent structures and heat transfer in multiple impinging jet arrays. Journal of Fluid Mechanics. 2006; 546:255-284.\\

[20] Geers, F. G., Tummers, M. J. \& Hanjalic, G. K. 2004 Experimental investigation of impinging jet arrays. Experiments in Fluids 36, 946-958.\\

[21] Lianmin Huang, Mohamed S. El-Genk, Heat transfer of an impinging jet on a flat surface, International Journal of Heat and Mass Transfer, Volume 37, Issue 13, 1994, 1915-1923. \\

[22] Hall JW, Ewing D. On the dynamics of the large-scale structures in round impinging jets. Journal of Fluid Mechanics. 2006; 555:439-458.\\

[23] Ruith MR, Chen P, Meiburg E, Maxworthy T. Three-dimensional vortex breakdown in swirling jets and wakes: direct numerical simulation. Journal of Fluid Mechanics. 2003; 486:331-378.\\

[24] Leibovich, Sidney. “The structure of vortex breakdown.” Annual Review of Fluid Mechanics 10 (1978): 221-246.\\

[25] Ianiro A, Lynch KP, Violato D, Cardone G, Scarano F. Three-dimensional organization and dynamics of vortices in multichannel swirling jets. Journal of Fluid Mechanics. 2018; 843:180-210.\\

[26] Mark J. Tummers, Jeroen Jacobse, Sebastiaan G.J. Voorbrood, Turbulent flow in the near field of a round impinging jet, Int J Heat and Mass Transfer, 54, Issues 23–24, 2011, 4939-4948.\\

[27] Oberleithner K, Sieber M, Nayeri CN, et al. Three-dimensional coherent structures in a swirling jet undergoing vortex breakdown: stability analysis and empirical mode construction. Journal of Fluid Mechanics. 2011; 679:383-414. \\

[28] Premchand V. Chandra, Pratikash P. Panda, Pradip Dutta, Flow and heat transfer characteristics of an impinging jet augmented with swirl, Int Comm Heat and Mass Transfer, 149, 2023, 107121 \\

\end{document}